\documentclass{article}
\begin{document}


\def\IR{\mathchoice{ \hbox{${\rm I}\!{\rm R}$} }
                   { \hbox{${\rm I}\!{\rm R}$} }
                   { \hbox{$ \scriptstyle  {\rm I}\!{\rm R}$} }
                   { \hbox{$ \scriptscriptstyle  {\rm I}\!{\rm R}$}}}

\def\IN{\mathchoice{ \hbox{${\rm I}\!{\rm N}$} }
                   { \hbox{${\rm I}\!{\rm N}$} }
                   { \hbox{$ \scriptstyle  {\rm I}\!{\rm N}$} }
                   { \hbox{$ \scriptscriptstyle  {\rm I}\!{\rm N}$}}}

\def\1I{\mathchoice{ \hbox{${\rm 1}\hspace{-2.3pt}{\rm l}$} }
                   { \hbox{${\rm 1}\hspace{-2.3pt}{\rm l}$} }
                   { \hbox{$ \scriptstyle  {\rm I}\!{\rm N}$} }
                   { \hbox{$ \scriptscriptstyle  {\rm I}\!{\rm N}$}}}

\def\IC{\mathchoice{ \hbox{${\rm l}\!\!\!{\rm C}$} }
                   { \hbox{${\rm l}\!\!\!{\rm C}$} }
                   { \hbox{$ \scriptstyle  {\rm l}\!\!\!\!\:{\rm C}$} }
                   { \hbox{$ \scriptscriptstyle {\rm l}\!\!\!\!\:{\rm C}$} } }

\def\IP{\mathchoice{ \hbox{${\rm I}\!{\rm P}$} }
                   { \hbox{${\rm I}\!{\rm P}$} }
                   { \hbox{$ \scriptstyle  {\rm I}\!{\rm P}$} }
                   { \hbox{$ \scriptscriptstyle  {\rm I}\!{\rm P}$}}}

\def\IE{\mathchoice{ \hbox{${\rm I}\!{\rm E}$} }
                   { \hbox{${\rm I}\!{\rm E}$} }
                   { \hbox{$ \scriptstyle  {\rm I}\!{\rm E}$} }
                   { \hbox{$ \scriptscriptstyle  {\rm I}\!{\rm E}$}}}

\def\kasten#1{\mathop{\mkern0.5\thinmuskip
                      \vbox{\hrule
                            \hbox{\vrule
                                  \hskip#1
                                  \vrule height#1 width 0pt
                                  \vrule}%
                            \hrule}%
                      \mkern0.5\thinmuskip}}

\def\qed{\mathchoice{\kasten{8pt}}%
                       {\kasten{6pt}}%
                       {\kasten{4pt}}%
                       {\kasten{3pt}}}%

\renewcommand{\labelenumi}{{\rm(\roman{enumi})}}

\def\epsi{\varepsilon}

\newcounter{mathe}
\newenvironment{mathe}[1]
{ \par\bigskip\refstepcounter{mathe}\noindent\textbf{#1 \arabic{mathe}.
 }\mdseries\ }
{\par\bigskip\mdseries}


\title{\bf Semi-classical motion of  dressed  electrons}

\author{Stefan Teufel and Herbert Spohn\\ \\
\em Zentrum Mathematik, Technische Universit\"at M\"unchen,\\
\em 80290 M\"unchen, Germany\\
email: teufel@ma.tum.de, spohn@ma.tum.de}
\date{October 5, 2000}
\maketitle
\bigskip

\begin{abstract}
We consider an electron coupled to the quantized radiation field and subject to a slowly 
varying electrostatic potential. We establish that over sufficiently long times radiation 
effects are negligible and the dressed electron is governed by an effective one-particle 
Hamiltonian. In the proof only a few generic properties of the full Pauli-Fierz 
Hamiltonian $H_{\rm PF}$ 
enter. Most importantly, $H_{\rm PF}$ must have an isolated ground state band for $|p|<
p_{\rm c}\leq \infty$ with $p$ the total momentum and $p_{\rm c}$ indicating that the
ground state band may terminate. This structure demands a local approximation theorem,
in the sense that the one-particle approximation holds until the semi-classical 
dynamics violates $|p|<p_{\rm c}$. Within this framework we prove an abstract Hilbert space 
theorem which uses no additional information on the Hamiltonian away from the band of interest. 
Our result is applicable to other time-dependent semi-classical problems.
We discuss semi-classical distributions for the effective one-particle dynamics
and show how they can be translated to the full dynamics by our results.
\end{abstract}

\newpage

\section{Introduction}

Electrons, protons, and other elementary charged particles move in essence
along classical orbits provided the external potentials have a slow variation. E.g.\ in
accelerators it is safe to compute the orbits by means of classical mechanics.
As a somewhat crude physical picture one imagines that the electron dressed with its 
photon cloud at any given instant $t$ is in its state of lowest energy consistent with the current
total momentum $p(t)$. During the time span $\delta t$ the external forces change the 
momentum of the electron followed by a rapid adjustment of the photon cloud resulting in
the new total momentum $p(t+\delta t)$. In this picture two physical mechanisms are interlaced.
Since the forces are weak, the acceleration is small and radiative losses can
be ignored. The effective dynamics of the dressed electron is conservative within a good
approximation. In addition the total momentum $p(t)$ has a slow variation and can be regarded 
as a semi-classical variable with respect to the effective Hamiltonian of the ground
state band (lowest energy shell). The goal of our paper is to establish the validity of
the physical picture.

To set the scene, let us introduce the quantum Hamiltonian under consideration. In fact, 
as will be explained below, our technique is fairly general and uses only a few generic 
properties of the Hamiltonian. Still, it is useful to have a specific example in mind.
We consider a free electron coupled to the quantized Maxwell field. The Hilbert space of
states for the electron is $\mathcal{H}_{\rm el}=L^2(\IR^3)$ and its time evolution is governed by 
the Hamiltonian 
$-\frac{1}{2m}\Delta$. $m$ is the mass of the electron and we have set $\hbar=1$. 
For the photon field we introduce the Fock space $\mathcal{F}_{\rm b} = \mathcal{H}_{\rm f}$
over the one-particle space $L^2(\IR^3)\otimes \IC^2$, i.e.\ $\mathcal{F}_{\rm b} =
\bigoplus_{n=0}^\infty S_n(L^2(\IR^3)\otimes \IC^2)^{\otimes n}$ and $S_n$ the
symmetrizer. Thus a state $\phi\in \mathcal{F}_{\rm b}$ is a sequence of vectors 
$\{\phi^{(0)}, \phi^{(1)}, \ldots\}$ with $\phi^{(n)}\in 
S_n (L^2(\IR^3)\otimes \IC^2)^{\otimes n}$ such that $\|\phi\|^2 = \sum_{n=0}^\infty 
\|\phi^{(n)}\|^2<\infty$. On $\mathcal{F}_{\rm b}$ we define the two component Bose
field with annihilation operators $a(k,\lambda)$, where $k\in \IR^3$ stands for the wave 
number and $\lambda =1,2$ for the helicity of the photon. The fields satisfy the canonical
commutation relations $[a(k,\lambda), a^*(k',\lambda')] = \delta (k-k') 
\delta_{\lambda\lambda'}$, $[a(k,\lambda), a(k',\lambda')] = 0 = 
[a^*(k,\lambda), a^*(k',\lambda')]$.
The Hamiltonian of the free photon field is the given by 
\begin{equation} \label{PFHf}
H_{\rm f} = \sum_{\lambda=1,2} \int d^3k\, \omega(k)a^*(k,\lambda)a(k,\lambda)
\end{equation}
with dispersion relation $\omega(k)=|k|$, the velocity of light $c=1$. The electron and 
the photon field are minimally coupled through the transverse vector potential $A(x)$.
To assure transversality we introduce the standard dreibein $e_1(k)$, $e_2(k)$, $k/|k|$.
Then 
\begin{equation}
A(x) = (2\pi)^{-3/2}  \sum_{\lambda=1,2} \int d^3k\, \frac{1}{\sqrt{2\omega(k)}}
 e_\lambda(k) \left(e^{ik\cdot x}a(k,\lambda) + e^{-ik\cdot x} a^*(k,\lambda)\right)\,.
\end{equation} 
$A$ is an operator valued distribution. To make it an unbounded operator we smoothen over the form factor $\rho$ as
\begin{equation}
A_\rho(x) = \int d^3x'\, \rho(x-x')A(x')
\end{equation}
and assume that $\rho$ is radial, smooth, of rapid decrease, and normalized as
 $\int \,d^3x\,\rho(x) =1$. In the corresponding classical Hamiltonian $\rho$ would be
 the rigid charge distribution.
With all these preparations we can introduce the 
Pauli-Fierz operator of a free electron as
\begin{equation} \label{PFH}
H_0 = \frac{1}{2m}\big( -i\nabla \otimes {\bf 1} - eA_\rho(x)\big)^2 + {\bf 1} \otimes H_{\rm f}
\end{equation}
acting on $\mathcal{H} = \mathcal{H}_{\rm el}\otimes \mathcal{H}_{\rm f}$. $e$ is the charge
of the electron. In (\ref{PFH}) $x$ denotes the position operator of the electron
on $L^2(\IR^3)$. $H_0$ is invariant under translations jointly of the electron and
the photons. Thus the total momentum 
\begin{equation} \label{TM}
p = p_{\rm el} \otimes {\bf 1} + {\bf 1}\otimes p_{\rm f}\,,\qquad
p_{\rm f} =  \sum_{\lambda=1,2} \int d^3k\quad k \,a^*(k,\lambda)a(k,\lambda)\,,
\end{equation}
is conserved, $[H_0,p]=0$. This can be seen more directly by rewriting
(\ref{PFH}) in momentum representation as
\begin{equation} 
H_0 =  \frac{1}{2m}\big( p_{\rm el} \otimes {\bf 1} - eA_\rho(i\nabla_{p_{\rm el}})\big)^2 
+ {\bf 1} \otimes H_{\rm f}\,
\end{equation}
and then going to  a representation in which $p$ is diagonal.
This is achieved through the unitary transformation $T$ defined by
\begin{equation}
(T\psi)^{(n)} (p,k_1,\ldots,k_n) = \psi^{(n)}(p-\sum_{i=1}^n k_i,k_1,\ldots,k_n)\,,
\end{equation}
where $\psi^{(n)}$ is the $n$-particle sector component of $\psi$ in electron momentum 
representation.
The transformed Hamiltonian $T^{-1}H_0T$, again denoted by $H_0$, takes the form
\begin{equation} \label{PFHp}
H_0 =  \frac{1}{2m}\big( p - p_{\rm f}  - eA_\rho(0)\big)^2 
+ H_{\rm f}\,.
\end{equation}
We decompose $\mathcal H$ and $H_0$ on the spectrum of $p$,
\begin{equation}\label{decomp}
{\mathcal H}  =  \int^\oplus_{\IR^3}d^3p\, {\mathcal H}_p \,,\qquad
H_0  =  \int^\oplus_{\IR^3}d^3p\, H_0(p)\,.
\end{equation}
The spaces  ${\mathcal H}_p$ are isomorphic to $\mathcal{H}_{\rm f}$ in a natural sense and
will be identified in the following. $H_0(p)$ is just $H_0$ acting on $\mathcal{H}_{\rm f}$
for a given value of $p$.

Physically one expects to have a dressed electron state for given momentum $p$, at least if
$|p|\leq p_{\rm c}$. In our semi-relativistic model states with $|p|\geq p_{\rm c}$ decay 
through Cherenkov radiation to lower momentum states. Thus provisionally we assume that 
there exists a $p_{\rm c}$ such that for every $p\in\Lambda_{\rm g} = 
\{p: |p|\leq p_{\rm c} \}$ $H_0(p)$ has a unique ground state with energy $E(p)$,
i.e.\
\begin{equation}
H_0(p)\psi_0(p) = E(p)\psi_0(p)
\end{equation}
has a  solution $\psi_0(p)\in \mathcal{H}_{\rm f}$ which is unique up to scalar multiples
 and $E(p) = {\rm inf \, spec}(H_0(p))$. 
We will discuss in Section 3 under which additional conditions on $\widehat \rho$ our
assumptions can be verified.

$E(p)$ is the ground state band energy. The corresponding projection is
$P_{\rm g} = \int_{\Lambda_{\rm g}}^\oplus d^3p\,P_0(p)$ where $P_0(p)$ denotes the
orthogonal projection onto the one-dimensional subspace spanned by
 $\psi_0(p)$. The states in Ran$P_{\rm g}$ are called dressed electron states.
More explicitly
\begin{equation} \label{RanPg}
{\rm Ran}\, P_{\rm g} = \left\{ \int_{\Lambda_{\rm g}}^\oplus d^3p\, \phi(p)\psi_0(p) \in 
\mathcal{H};\,\phi\in L^2(\Lambda_{\rm g})\right\}\,.
\end{equation}
On Ran$P_{\rm g}$ we have
\begin{equation}
e^{-iH_0t}\psi =\int_{\Lambda_{\rm g}}d^3p\,\left( e^{-iE(p)t}\phi(p)\right)\psi_0(p)\,.
\end{equation}
Thus, if initially in Ran$P_{\rm g}$, the dressed electron propagates like a single quantum 
particle  with dispersion relation $E(p)$, which is generated through the interaction with the
photons. In particular, it stays in the dressed electron
subspace for all times.

As already remarked, the motion of an electron is modified through external electro-magnetic
potentials. In general, they have a slow variation on the scale set by the Compton 
wave length. Thus we add to $H_0$ the external potential $V(\epsi x)$ (and possibly also
an external vector potential $A_{\rm ex}(\epsi x)$). $\epsi$ is a small dimensionless parameter
which controls the variation of $V$. The external forces break the translation invariance
of $H_0$ and the total momentum is no longer conserved. A state $\psi$ initially in
Ran$P_{\rm g}$ will no longer remain so under the time evolution generated by the full 
Hamiltonian $H$, which reflects that an accelerated charge looses energy through
radiation.  Since the external forces are weak of order $\epsi$, we can expect radiation 
losses to be negligible. More precisely, the acceleration is of order $\epsi$ and by 
Larmor's formula the energy radiated over the time span $\tau$ is $\epsi^2\tau$.
Thus the relevant time scale is of order $\epsi^{-1}$. On that time scale the radiation loss 
is of order $\epsi$, whereas the cumulative effect of the forces is of order $1$ and the
electron moves on the scale set by the potential $V$. If the initial $\psi\in$ Ran$P_{\rm g}$,
the dressed electron should still be governed by an effective 
one-particle Hamiltonian, which is obtained from the dispersion relation $E(p)$ through
the Peierls substitution
\begin{equation} \label{H1}
H_1 = E\left(p-eA_{\rm ex}(\epsi x)\right) + V(\epsi x)\,,
\end{equation}
and
\begin{equation} \label{H1evo}
e^{-iHt}\psi = \int_{\Lambda_{\rm g}}d^3p\,\left( e^{-iH_1t}\phi(p)\right)\psi_0(p)
+O(\epsi)
\end{equation}
for $0\leq t\leq \epsi^{-1}T$ with some suitable macroscopic time $T$. We will
establish (\ref{H1evo}) under some additional assumptions on $H_0$ and for
$A_{\rm ex} = 0$.

In  (\ref{H1evo}) it is crucial to assume that $\psi\in$ Ran$P_{\rm g}$. For a general 
$\psi\in \mathcal{H}$ one expects that, in essence on a time scale of order $1$,
$\psi$ splits into outgoing radiation and a piece which is approximately in Ran$P_{\rm g}$.
This second piece is then governed by the approximation  (\ref{H1evo}). To prove that this
really happens is a problem of scattering theory for free electrons and outside the scope of
our present investigation. To our knowledge this is an unsolved and very challenging problem. 

There are several difficulties with the picture proposed in  (\ref{H1evo}).
 Firstly the Pauli-Fierz Hamiltonian
is infrared divergent. For $p\not=0$, the photon cloud has an infinite number of photons,
though of finite total energy. The physical ground state at fixed $p\not=0$ does not lie in
Fock space. Even if we introduce a suitable infrared cut-off by assuming
that $\widehat \rho(k)\to 0$ as $k\to 0$, with $\widehat \rho$ the Fourier transform of
$\rho$, $E(p)$ is not separated from the rest of the spectrum of $H_0(p)$.
At present, we have no technique available to control (\ref{H1evo}) in case
$H_0(p)$ has no spectral gap. To overcome both difficulties we are forced to give the photons a 
small mass, which means to set $\omega(k) = (m_{\rm ph}^2 + k^2)^{1/2}$.
We emphasize that the radiation losses are not affected by this cut-off.

As a second difficulty, which has been accounted for already, we observe
 that $E(p)$ will cease to exist beyond
a certain critical value $p_{\rm c}$, i.e.\ for $|p|\geq p_{\rm c}$. This is most easily
understood by considering the uncoupled Hamiltonian $H_0(p)$ at $e=0$. It has absolutely 
continuous spectrum  and the only eigenvalue $E(p)= \frac{1}{2m_{\rm el}} p^2$.
This eigenvalue is isolated and below the continuum edge 
for $|p|<m_{\rm el}$ ($m_{\rm ph}$ small)
and is embedded in the continuum for $|p|>m_{\rm el}$. 
For small coupling the embedded eigenvalue
should dissolve and $E(p)$ exists only for $|p|<p_{\rm c} \cong m_{\rm el}$.

We could avoid the bounded extension of $E(p)$ through a suitable modification of the
boson dispersion $\omega(k)$ and/or the electron dispersion $p^2/2m$, cf.\ Section 2.3.
 However, the termination of bands is a fairly
generic phenomenon. Thus, viewed in a more general setting, we have a classical Hamiltonian 
\begin{equation}
H_{\rm cl} (q,p) = E(p) + V(q)
\end{equation}
corresponding to (\ref{H1}) on phase space $\Gamma=\IR^3\times \Lambda_{\rm g}$.
The solution flow does not exist globally in time and for given initial conditions
$(q,p)$ there is a first time $T$ (including $T=\infty$) when the solution trajectory hits 
the boundary of $\Gamma$. This means we have to control the approximation (\ref{H1evo})
up to the time when substantial parts of the wave packet ``leave'' the allowed phase space.
Beyond that time new physical phenomena appear not accounted for by (\ref{H1evo}).
We believe that one of our main achievements is to develop a technical machinery
which allows for such local approximations.

It turns out that the validity of (\ref{H1evo}) relies only on a few rather general facts.
Therefore we decided to prove (\ref{H1evo}) as an abstract operator problem. The basic
assumptions are the decomposition (\ref{decomp}) and a non-degenerate energy band
of possibly finite extension separated by a gap from the rest of the spectrum.
We mention three widely studied physical systems which posses this abstract mathematical
structure, but are, at first sight, very different from the dressed
electron.

\noindent (i) For an electron moving in a periodic crystal potential, 
$p$ becomes the quasi-momentum
and the energy band is one particular Bloch band. Bloch bands are discrete but may cross.
The small parameter arises, as for Pauli-Fierz, through a, on the scale of the lattice
spacing, slowly varying external potential. In \cite{HST} we studied the semi-classical 
motion in periodic potentials for Bloch bands that are isolated over the whole 
Brillouin zone. While we used a similar approach, the present paper is a substantial
improvement in two respects. The approximation is local, no isolated bands have to be assumed, 
and we consistently avoid using any information on $H(p)$ away from the band of interest.

\noindent (ii) Electrons are lighter than nuclei by a factor of $2\cdot 10^3$--$5\cdot 10^5$,
which is the starting point of the Born-Oppenheimer approximation for molecular dynamics. 
$p$ is now replaced by
$\vec R$, the collection of nucleonic coordinates. $H(\vec R)$ is the electron 
Hamiltonian for fixed nuclei and the band structure arises from eigenvalues of
 $H(\vec R)$. They may cross or dive into the continuum. If we include the kinetic energy
of the nuclei, the small parameter becomes $\epsi = (m_{\rm el}/m_{\rm nucleus})^{1/2}$
and, except for the interchange of $p$ and $\vec R$, the Born-Oppenheimer approximation
fits our framework. Since in the following only bounded
potentials will be covered---the kinetic energy is $p^2$ and plays the role of the potential---we 
postpone a discussion of the time-dependent 
Born-Oppenheimer approximation to a separate paper. 
For an analysis from the point of view of wave packet dynamics we refer to \cite{Hagedorn}.

\noindent (iii) The Dirac equation for a single particle has the electron and the positron
band. One studies the motion
of an electron, say, under slowly varying external potentials 
\cite{Bolte}, \cite{Spohn2}. As novel feature, the bands are doubly degenerate.
Presumably this would also be the case if we include the electron spin in (\ref{PFHp}) as
\[
H= \frac{1}{2m}\big( \sigma\cdot(p-p_{\rm f}-eA_\rho)\big)^2 + H_{\rm f}\,.
\]
In the semi-classical limit the internal degrees of freedom (degeneracy) remain
quantum mechanical and one has to approximate by matrix valued classical mechanics. 

Our paper is organized as follows.
The abstract setting will be explained in Section 2. In this framework, almost by necessity, the
theorems are stated as uniform convergence of certain unitary groups and of the 
corresponding time dependent semi-classical observables in the Heisenberg picture. 
The reader may worry, as we did, whether such convergence results imply the
semi-classical approximation of quantities of physical interest, like position and momentum 
distributions. In fact, they do under very mild assumptions on the initial
wave function. We could not find a coverage of sufficient generality in the literature
and therefore discuss semi-classical distributions in Section 6. Our main theorems are stated
in Section 2 with proofs given in Sections 3 to 5.

\section{General setting and main results}

\subsection{General setting}

As the ``momentum space'' $M$ we take either $\IR^d$, $d\in\IN$, in which case $dp$ will denote
Lebesgue measure on $\IR^d$, or a flat $d$-torus. In the latter case we take $dp$ to denote
normalized Lebesgue measure on the torus. In the following $\mathcal{H}_{\rm f}$ denotes
any separable 
Hilbert space, although the notation should remind one of the Hilbert space for the Bose field
in case of our main application.

Let $H_0$ be a self-adjoint operator on $D(H_0)\subset
\mathcal H = L^2(M) \otimes \mathcal{H}_{\rm f}$ 
that can be decomposed on  $M$
as
\begin{equation}
H_0 = \int^\oplus_M dp\,H_0(p)\,,
\end{equation}
where $\{H_0(p), p\in M\}$ is a family of self-adjoint operators with a common domain 
$\mathcal{D}_0\subset \mathcal{H}_{\rm f}$.
We assume that the map $p\mapsto H_0(p)$ is differentiable in the  sense that for all $p\in M$
and $ j=1,\ldots, d$ the limit
\[
\big(\partial_{p_j}H_0(p)\big) (H_0(p) - i)^{-1} := \lim_{h\to 0} \frac{H_0(p+he_j) - H_0(p)}{h} (H_0(p) - i)^{-1}
\]
exists in the norm of bounded operators on $\mathcal{H}_{\rm f}$. This defines, in particular, 
$(\nabla_p H_0)(p)$ as a self-adjoint operator on $\mathcal{D}_0^d$.

For $p$ in some compact and convex $\Lambda_{\rm g}\subseteq M$ with non-empty interior let $H_0(p)$ have an 
isolated non-degenerate  eigenvalue
$E(p)$, i.e.\ 
\[
H_0(p)\psi_0(p)= E(p)\psi_0(p)
\]
with $\psi_0(p)\in \mathcal{H}_{\rm f}$ and
dist$\big(E(p),\sigma(H_0(p))\setminus E(p)\big)>0$. We assume that $E(\cdot)\in C^\infty(\Lambda_{\rm g}, \IR)$. As before, $P_0(p)$ denotes the rank-one projection 
onto $\psi_0(p)$ and $P_{\rm g}$ is defined as the orthogonal projection on 
Ran$P_{\rm g}$ given by (\ref{RanPg}).
The eigenfunctions $\psi_0(p)$, $p\in\Lambda_{\rm g}$, are defined only up to a
$p$-dependent phase factor. One can choose this phase factor such that $\psi_0(\cdot)\in 
C^2(\Lambda_{\rm g},\mathcal{H}_{\rm f})$ (cf.\ Lemma \ref{HdiagH1}),
and we assume such a choice in the following.
On $\Lambda_{\rm g}$ we require $H_0(\cdot)$ to be twice continuously differentiable in the same sense
as above, i.e.\  for all $p\in \Lambda_{\rm g}$ and $j,k\in 1,\ldots, d$ the limit
\[
\big(\partial_{p_k}\partial_{p_j}H_0(p)\big) (H_0(p) - i)^{-1} := 
\lim_{h\to 0} \frac{(\partial_{p_j} H_0)(p+he_k) -(\partial_{p_j} H_0)(p)}{h} (H_0(p) - i)^{-1}
\]
exists in the norm of bounded operators on $\mathcal{H}_{\rm f}$ and depends continuously
on $p$ in the same topology.

Let the potential $V:\IR^d\to\IR$ be such that 
$V(x) = \int dk\, e^{ik\cdot x}\, \widehat V(k)$, where 
$\int dk\, |k|^n |\widehat V(k)|<\infty$ for all $n\in\IN_0$. 
This guarantees, in particular, that $V$ and all its partial derivatives are in
$C^\infty(\IR^d)$ and vanish at infinity.
$V^\epsi := V(i\epsi\nabla_p)$ is a bounded self-adjoint operator on $L^2(M)$ and
the full Hamiltonian 
\begin{equation}
H = H_0 + V^\epsi\otimes {\bf 1}_{\mathcal{H}_{\rm f}}
\end{equation}
is self-adjoint on $D(H_0)$. 

We define
the corresponding one-particle Hamiltonian on $L^2(M)$ by
\begin{equation}
H_1 = E(p) + V^\epsi\,.
\end{equation}
Note that $E(p)$ is a priori only defined for $p\in \Lambda_{\rm g}$. 
To make things simple, 
we continue $E(p)$ to a smooth and compactly supported but otherwise arbitrary function 
on $M$.
Since we will be interested only in the behavior for $p\in \Lambda_{\rm g}$, we use 
this global function
to define $H_1$.
The corresponding unitary groups are denoted by $U(t) = e^{-iHt}$ and $U_1(t) = e^{-iH_1t}$.

Our goal is to show that, in a suitable sense,
\begin{equation} \label{comp}
 \big( U\left(t/\epsi\right) - U_1\left(t/\epsi
\right)\big) P_{\rm g}=O(\epsi)
\end{equation}
for macroscopic times $t<T<\infty$, where $T$ depends only on $V$ and the initial momenta.

But before we can state the precise result, we have to make sense of $H_1$ acting on
Ran$P_{\rm g}$. According to (\ref{RanPg}) we define the map $\mathcal{U}: 
{\rm Ran}P_{\rm g} \to L^2(M)$
\begin{equation}
\mathcal{U}(\phi\psi_0) = \phi\,.
\end{equation}
Its adjoint $\mathcal{U}^*: L^2(M) \to  {\rm Ran}P_{\rm g}$ is given by
\begin{equation}
\mathcal{U}^* \phi = \int^\oplus dp\, \1I_{\Lambda_{\rm g}} (p) \phi(p)\psi_0(p)\,,
\end{equation}
where here and in the following $\1I_A$ denotes the characteristic function of a set $A$.
Clearly $\mathcal{U}$ is an isometry and  $\mathcal{U}^*\mathcal{U} = \bf 1$ on Ran$P_{\rm g}$.

The effective dynamics cease to make sense, once the momentum of the particle leaves
$\Lambda_{\rm g}$. If that is not excluded by energy conservation, the
comparison (\ref{comp}) can hold only up to some finite time, which can be determined
from the classical dynamics generated by 
\[
H_{\rm cl}(x,p)= E(p) + V(x)
\] 
on phase space $\IR^d\times M$.
Let $\Lambda_{\rm i} \subset M$, 
the ``set of initial momenta'', and $\Lambda_{\rm m} \subset M$
some ``maximal  set of allowed momenta'', both be compact. 
For $\Lambda\subset M$ compact and $\delta\geq 0$ we let
\[
\Lambda+\delta :=
\left\{ p\in M: \inf_{k\in \Lambda} |p-k|\leq\delta\right\}
\]
be the corresponding $\delta$-enlarged set, which is again compact.
With
$\Phi_p^t:\IR^d\times M\to M$ denoting the momentum component of the classical flow, we define
for $\Lambda_{\rm i}+\delta \subset \Lambda_{\rm m}$ 
\[
T_{\rm m}^\delta (\Lambda_{\rm i}, \Lambda_{\rm m}) 
:= \sup_{t\geq 0}\,\left\{t\,:\, {\rm supp}\,\Big(\big(\1I_{\Lambda_{\rm i}}\circ \Phi_p^{-s}\big)
(x,\cdot)\Big)  
+\delta \subseteq \Lambda_{\rm m} \,\, \forall \, s\in[0,t],x\in \IR^d\right\}
\]
as the maximal time for which  the momentum of any (i.e.\ with any starting position) 
classical particle with initial momentum
in $\Lambda_{\rm i}$ stays within a $\delta$-margin
inside of $\Lambda_{\rm m}$.

The following lemma shows that the classical bound on the momentum is respected also by
the quantum dynamics in the
limit $\epsi\to 0$. Let $P_{\rm i} = \1I_{\Lambda_{\rm i}}(p)$ and  
$P_{\rm m} = \1I_{\Lambda_{\rm m}}(p)$.

\begin{mathe}{Lemma}\label{Impulserhaltung} 
Let $\Lambda_{\rm i}$, $\Lambda_{\rm m}\subset M$ be both compact
and $\Lambda_{\rm i}+\delta\subset\Lambda_{\rm m}$ for some $\delta>0$. 
For any $T<\infty$ with $T\leq T^\delta_{\rm m}(\Lambda_{\rm i},\Lambda_{\rm m})$ 
there is a $C<\infty$
such that for all $t\in[0,T]$
\begin{equation} \label{ipu}
\left\|\left( 1- P_{\rm m} \right) \,
 U_1\left(\frac{t}{\epsi}\right) 
P_{\rm i} 
\right\|_{\mathcal{L}(L^2(M))} \leq C\,\epsi^2\,.
\end{equation}
\end{mathe}
For the proof of Lemma \ref{Impulserhaltung} see Section 3.

\subsection{Main results}
 
In the following we will always assume
that $\Lambda_{\rm i} \subset \Lambda_{\rm g}$ and 
we can therefore identify
$P_{\rm i}$ with  
$\mathcal{U}^*P_{\rm i}\mathcal{U}P_{\rm g}$ to keep notation simple. 
I.e., $P_{\rm i}$ projects
on ``dressed electron states'' with momenta in $\Lambda_{\rm i}$.

\begin{mathe}{Theorem}\label{MT}
Let $\Lambda_{\rm i}$ be compact with $\Lambda_{\rm i}+\delta \subset \Lambda_{\rm g}$ for some
$\delta>0$. For any $T<\infty$ with $T\leq T^\delta_{\rm m}(\Lambda_{\rm i},\Lambda_{\rm g})$
there is a $C<\infty$
such that for all  $t\in[0,T]$
\begin{equation}\label{MTe}
\left\| \left(
U\left(\frac{t}{\epsi}\right) - \mathcal{U}^*U_1 \left(\frac{t}{\epsi}\right)
\mathcal{U}\right) P_{\rm i}\right\|_{\mathcal{L}(\mathcal{H})}\leq C\,\epsi \,.
\end{equation}
\end{mathe}
This means, in the case of our main application, 
that dressed electron states with initial momenta in $\Lambda_{\rm i}$
evolve according to the effective dynamics without radiation losses for times of order
$\epsi^{-1}$ as long as the momenta of the corresponding classical orbits
stay inside $\Lambda_{\rm g}$. Note, in particular, that the macroscopic time $T^\delta_{\rm m}$
depends only on $\Lambda_{\rm i}$, but not on $\epsi$.

The equivalence between the full dynamics and the effective one-particle 
dynamics on the isolated energy band
extends to the level of semi-classical or macroscopic observables.
On a suitable domain in $\mathcal{H}$ the macroscopic (Heisenberg) position 
operator is given
by
\[
x^\epsi(t) = U\left(-t/\epsi\right) \,\left(\epsi i\nabla_p\otimes{\bf 1}\right)
 \, U\left(t/\epsi\right)
\]
and its effective one-particle counterpart on $H^1(\IR^d)$ by
\[
x^\epsi_1(t) =
U_1\left(-t/\epsi\right) \,\epsi i\nabla_p \, U_1
\left(t/\epsi\right)\,.
\]

\begin{mathe}{Theorem} \label{effop}
Let $\Lambda_{\rm i}$ be compact with $\Lambda_{\rm i}+\delta \subset \Lambda_{\rm g}$ for some $\delta>0$. 
For every $T<\infty$ with $T\leq T^\delta_{\rm m}(\Lambda_{\rm i},\Lambda_{\rm g})$ there is a $C<\infty$
such that for all  $t\in[0,T]$
\begin{equation} \label{xeff}
\Big\| \Big( x^\epsi(t) - \mathcal{U}^*x^\epsi_1(t)\mathcal{U}\Big) 
P_{\rm i} \Big\|_{\mathcal{L(H)}} <\,C\,\epsi\,.
\end{equation}
\end{mathe}

 Theorem \ref{effop} is not a trivial corollary of Theorem 
\ref{MT} for several reasons.
It is a statement about unbounded operators, whose difference, however, is bounded.
More importantly, the difference of
the position operators does not vanish even at time $t=0$ because $[x^\epsi(0),P_{\rm g}]\not=0$.

More generally, consider classical symbols 
 $a\in C^\infty(\IR^d\times M)$ such that for
all multi-indices $\alpha$, $\beta$ there exists $C_{\alpha,\beta}<\infty$ with
\[
\sup_{x,p} \left| \partial^\alpha_x \partial^\beta_p a(x,p) \right| \leq C_{\alpha,\beta}\,.
\]
We use the notation of \cite{DS} and denote this  
set of symbols by $S^0_0(1)$.
For $a\in S^0_0(1)$ 
Weyl quantization leads to an operator $a^{\rm W,\epsi}\in \mathcal{L}(L^2(M))$
given by 
\begin{equation}\label{WeylDef}
\left( a^{\rm W, \epsi} \phi\right)(p) = (2\pi)^{-d} \int dx \,dk\,\,a\left(\epsi x, 
\frac{p+k}{2}\right)\, e^{-i(p-k)\cdot x}\,\phi(k) \,,
\end{equation}
with operator norm that is bounded uniformly in $\epsi$ (cf.\ Section 4 for details).

For $a\in S^0_0(1)$ let
\begin{equation} \label{adef}
a^\epsi(t) =  U\left(-t/\epsi\right) \,\left(
a^{\rm W, \epsi}\otimes{\bf 1}\right) \, U\left(t/\epsi\right)
\end{equation}
and 
\begin{equation}\label{a1def}
a^\epsi_1(t) =
U_1\left(-t/\epsi\right) \,a^{\rm W, \epsi} \, U_1
\left(t/\epsi\right)\,.
\end{equation}
\begin{mathe}{Theorem} \label{effop2}
Let $\Lambda_{\rm i}$ be compact with $\Lambda_{\rm i}+\delta \subset \Lambda_{\rm g}$ for some $\delta>0$ and $a\in S^0_0(1)\cap C_\infty(\IR^d\times M)$.
For every $T<\infty$ with $T\leq T^\delta_{\rm m}(\Lambda_{\rm i},\Lambda_{\rm g})$ there is a $C<\infty$
such that for all  $t\in[0,T]$
\begin{equation} \label{xeff2}
\Big\| \Big( a^\epsi(t) - \mathcal{U}^*a^\epsi_1(t)\mathcal{U}\Big) 
P_{\rm i} \Big\|_{\mathcal{L(H)}} <\,C\,\epsi\,.
\end{equation}
\end{mathe}

 $C_\infty$ denotes the set of continuous functions vanishing at infinity. Presumably
Theorem \ref{effop2} holds as well for $a\in S^0_0(1)$, however, the assumption $a\in C_\infty$
allows for a simpler approximation argument in the proof.

\subsection{The massive Nelson and Pauli-Fierz model}

It is of interest to see whether our abstract assumptions are in fact satisfied for physically,
at least semi-realistic models. The best studied case is the Nelson model \cite{Nelson},
where the coupling is to the position of the particle and the Maxwell field is replaced by a
scalar field. Switching immediately to the total momentum representation, cf.\ (\ref{PFHp}),
the Nelson Hamiltonian reads
\begin{equation} \label{NHp}
H_{\rm N}(p) = \frac{1}{2} (p-p_{\rm f})^2 + H_{\rm f} + (2\pi)^{-d/2} 
\int dk\, \frac{\widehat \rho(k)}{\sqrt{2\omega(k)}} 
\left( a(k) + a^*(-k)\right)\,,
\end{equation}
where instead of (\ref{PFHf}), (\ref{TM}), $a(k)$, $a^*(k)$ is a one-component Bose field 
over $\IR^d$. Again, $p$ is the total momentum and regarded as a parameter. If
$\int dk\, |\widehat\rho|^2(\omega^{-3} + \omega^{-1}) <\infty$, then $H_{\rm N}(p)$ is bounded from below
and self-adjoint with domain $D(H_{\rm f})$.

According to a result of Fr\"ohlich \cite{Froehlich}, under this condition with 
$\omega(k) = (m_{\rm ph}^2+k^2)^{1/2}$,  $m_{\rm ph}>0$ and in dimension $d=3$, 
$H_{\rm N}$  has an isolated, nondegenerate
ground state band for $|p|<p_{\rm c}$ with  $p_{\rm c}>\sqrt{3}-1$. 
 If in (\ref{NHp}) we replace the electron dispersion by $E_0(p) = (m_{\rm el}^2+p^2)^{1/2}$,
$m_{\rm el}>0$, still keeping $\omega(k) = (m_{\rm ph}^2+k^2)^{1/2}$,  $m_{\rm ph}>0$, then
at zero coupling the ground state lies strictly below the continuum edge for all $p$,
which means $p_{\rm c} = \infty$ for $\widehat \rho = 0$. As proved in \cite{Froehlich}
$p_{\rm c} =\infty$ persists to arbitrary  coupling strength.

A larger class of bosonic dispersion relations
is studied in \cite{Spohn}. For the particular case $\omega(k)=\omega_0>0$ and dimensions $d=1,2$
one has $p_{\rm c}=\infty$, whereas  for $d=3$ and small coupling indeed $p_{\rm c}<\infty$ \cite{M}.

Fr\"ohlich encloses the Bose field in a periodic box and proves that as the box size goes to infinity,
there is still spectrum of $H_{\rm N}(p)$ strictly 
 below the continuum edge, provided $|p|<p_{\rm c}$ which has to be estimated through a separate
argument.
Thus $H_{\rm N}(p)$ has a ground state for  $|p|<p_{\rm c}$. To establish uniqueness he 
uses that $e^{-tH_{\rm N}(p)}$ in Fock space representation is positivity improving, except for a 
small error which goes to zero as $t\to\infty$. From analytic perturbation theory one concludes
then that the ground state band is isolated and analytic in $p$, provided  $|p|<p_{\rm c}$.
Hence, the massive Nelson model satisfies the assumptions underlying 
Theorem \ref{MT},  \ref{effop} and  \ref{effop2}.

For the Pauli-Fierz model (\ref{PFHp}) the existence part of the proof goes through without 
essential changes.
Unfortunately, the positivity part fails and  no argument is known to ensure
 the uniqueness of the ground state for $|p|<p_{\rm c}$.
Since for $e=0$ and small $p$ the ground state band is isolated, it must remain so for small $e$
by perturbation theory, however, with the undesirable feature that smallness depends now on 
$m_{\rm ph}$.
Even for the massive Pauli-Fierz model a 
general proof of uniqueness would be most welcome.

\section{Preliminaries  from semi-classical analysis}

Although Theorem \ref{MT} compares a full quantum evolution to an effective quantum evolution
and says nothing about semi-classics, its proof makes use of Lemma \ref{Impulserhaltung}.
To prove Lemma \ref{Impulserhaltung} we need some tools from semi-classical analysis, which
are introduced and applied to the one-particle Hamiltonian $H_1$ in this section.
Furthermore, in Section 6 we will consider semi-classical distributions based
on the notions and results discussed here.

Since $E$ and $V$ are both smooth bounded functions, we can apply standard results
of semi-classics to the Hamiltonian 
\[
H_1(p,i\epsi\nabla_p) = E(p) + V(i\epsi\nabla_p)
\]
acting on $L^2(M)$, where the roles of momentum and position are 
exchanged and the role
of $\hbar$ is taken by $\epsi$.
In the following we will simply ignore this difference and leave the necessary
changes as compared to the standard case to the reader. Note, however, that
the change in sign, $i\nabla$ instead of $-i\nabla$, is canceled
by the fact that $p' = q$ and $q'=-p$ is the canonical transformation interchanging
$q$ and $p$.

We will consider classical symbols  $a\in S^0_0(1)$. As was stated in Section 2, 
Weyl quantization of functions in   $S^0_0(1)$ leads to bounded operators. 
The following result sharpens this statement.

\begin{mathe}{Proposition}[Calderon-Vaillancourt]\label{CV}
There is a $C<\infty$ and a finite $n\in\IN$ such that
for all  $a\in S^0_0(1)$ and $\epsi\in[0,1]$
\begin{equation}
\big\| a^{\rm W,\epsi} \big\|_{\mathcal{L}(L^2( M))} \leq
C\, \sup_{|\alpha|\leq n, |\beta|\leq n, (x,p)\in\IR^d\times M} \big|
\partial^\alpha_x\partial^\beta_p \,a(x,p)\big|\,.
\end{equation} 
\end{mathe}
For a proof see \cite{DS}, Theorem 7.11. The statement there is slightly different, 
but their proof implies our Proposition \ref{CV}. Note, in particular, that
the constants $n$ and $C$ in Proposition \ref{CV} depend on the dimension
$d$ of configuration space.

The so called product rule, presented in the following Proposition \ref{basicprod},
is at the basis of all semi-classical analysis we will apply. Given Proposition
\ref{CV}, however, its proof is mainly a matter of calculation. 
For details see \cite{DS}, Proposition 7.7 and Theorem 7.9.

\begin{mathe}{Proposition}[Product rule] \label{basicprod}
Let $a,b\in S_0^0(1)$. Then for all $n\in \IN_0$ there is a $d_n<\infty$ such that
\begin{equation} \label{prodeq}
\Big\| a^{\rm W,\epsi} b^{\rm W,\epsi} - (\sum_{k=0}^n \epsi^k\, c_k )^{\rm W,\epsi} \Big\|_{\mathcal{L}(L^2( M))} 
\leq \,d_n\,\epsi^{n+1}\,,
\end{equation}
with
\begin{equation}
c_k(x,p) = \left(\frac{i}{2}\right)^k \sum_{|\alpha|+|\beta|=k}
\frac{(-1)^{|\beta|}}{|\alpha|!|\beta|!}\Big( \,
(\partial_x^\beta \partial_p^\alpha a)
(\partial_p^\beta \partial_x^\alpha b) \Big)(x,p)\,.
\end{equation}
\end{mathe}
We state two simple facts that are immediate consequences of the product rule as
\begin{mathe}{Corollary} \label{product}
Let $a,b \in S_0^0(1)$. Then 
\begin{enumerate}
\item
\[
\Big\| a^{\rm W,\epsi}b^{\rm W,\epsi} - (ab)^{\rm W,\epsi} \Big\|_{\mathcal{L}(L^2( M))} = O(\epsi)\,.
\]
\item If, in addition, ${\rm supp}(a)\cap {\rm supp}(b) = \emptyset$, 
then for any $n\in \IN$
\[
\Big\| a^{\rm W,\epsi} b^{\rm W,\epsi} \Big\|_{\mathcal{L}(L^2( M))}=O(\epsi^n)\,.
\]
\end{enumerate}
\end{mathe}
\noindent{\bf Proof.} \quad(i) is just Proposition \ref{basicprod} for $n=0$. 
 (ii) holds since in this case $c_k=0$ in (\ref{prodeq}) for all $k\in\IN_0$.

\hfill $\qed$

Here and in the following $O(\epsi^n)$ means that an expression, or its appropriate norm,
is bounded by a constant times $\epsi^n$ for sufficiently small $\epsi$.  

The crucial ingredient to our semi-classical analysis is the following first-order
version of a Theorem going back to Egorov \cite{Egorov} (cf.\ also \cite{Robert}, Th\'eor\`eme
IV-10), 
which is also a direct consequence of the product rule. 

\begin{mathe}{Proposition}[Egorov's Theorem] \label{Egorov}
Let $a\in S^0_0(1)$ and $0\leq T<\infty$. 
There is a $C<\infty$ such that for all $t\in [-T,T]$
\[
\left\|  U_1\left(-\frac{t}{\epsi}\right) a^{\rm W,\epsi}
U_1\left(\frac{t}{\epsi}\right) - \left( a\circ \Phi^t \right)^{\rm W,\epsi}
\right\|_{\mathcal{L}(L^2(M))} \leq C\epsi^2\,.
\]
\end{mathe}

\noindent {\bf Proof.} First note that for $a\in S^0_0(1)$ we have that 
$a\circ\Phi^t\in S^0_0(1)$ for all finite $t$, since the
Hamiltonian vector field is smooth, uniformly bounded and all its partial derivatives
are uniformly bounded.
Moreover, $a\circ\Phi^t$ and all its partial derivatives are each bounded uniformly for
$t\in[-T,T]$. Therefore $\|(a\circ\Phi^t)^{\rm W,\epsi}\|_{\mathcal{L}(L^2( M))}$ is
bounded uniformly for $t\in[-T,T]$ by Proposition \ref{CV}.

Writing
\begin{eqnarray} \lefteqn{
 U_1\left(-\frac{t}{\epsi}\right) a^{\rm W,\epsi}
U_1\left(\frac{t}{\epsi}\right) - \left( a\circ \Phi^t \right)^{\rm W,\epsi}
=} \nonumber\\ &= &
\int_0^t ds\,\frac{d}{ds} \left(  U_1\left(-\frac{s}{\epsi}\right)
(a\circ\Phi^{t-s})^{\rm W,\epsi}  U_1\left(\frac{s}{\epsi}\right)
\right)\,,
\end{eqnarray}
one is led to consider
\begin{eqnarray} \label{egoeq}
\lefteqn{
 \frac{d}{ds} \, U_1\left(-\frac{s}{\epsi}\right)
(a\circ\Phi^{t-s})^{\rm W,\epsi}  U_1\left(\frac{s}{\epsi}\right) =}
\\
&=& U_1\left(-\frac{s}{\epsi}\right)
\left(
\frac{i}{\epsi} \left[ H_{\rm cl}^{\rm W,\epsi}, (a\circ\Phi^{t-s})^{\rm W,\epsi} \right] -
\left\{ H_{\rm cl} , (a\circ\Phi^{t-s}) \right\}^{\rm W,\epsi}
\right)
 U_1\left(\frac{s}{\epsi}\right)\,,\nonumber
\end{eqnarray}
where $\{\cdot,\cdot\}$ denotes the Poisson bracket and we used that 
$H_1= H_{\rm cl}^{\rm W,\epsi}$.
From the product rule one computes that for arbitrary $a,b\in S_0^0(1)$
\begin{equation}\label{commu}
\frac{i}{\epsi}\big[a^{\rm W,\epsi}, b^{\rm W,\epsi}\big] = 
\frac{i}{\epsi}\big( a^{\rm W,\epsi} b^{\rm W,\epsi} -  b^{\rm W,\epsi} a^{\rm W,\epsi} \big)
= \big\{a,b\big\}^{\rm W,\epsi} + O(\epsi^2)\,,
\end{equation}
which implies that (\ref{egoeq}) is $O(\epsi^2)$ for fixed $t-s$.
However, one can easily convince oneself that this bound is uniform for $t-s$ in any bounded
interval, since the semi-norm used in Proposition \ref{CV} is bounded uniformly for the $c_2$-terms
appearing in a derivation of Equation (\ref{commu}) using the product rule.

\hfill $\qed$\medskip

To have a natural way for extending functions of $x$ or $p$ alone to functions
on phase space, we introduce the projections $\pi_x: \IR^d\times M\to  \IR^d$
and $\pi_p: \IR^d\times M\to  M$ as $\pi_x (x,p) = x$ and $\pi_p (x,p) = p$. 
We are now ready to prove Lemma  \ref{Impulserhaltung}.
\medskip

\noindent {\bf Proof} (of Lemma  \ref{Impulserhaltung}). \quad
In order to regularize $\1I_{\Lambda_{\rm i}}$ and  $\1I_{\Lambda_{\rm m}}$
we pick some $f_{\rm i}\in C^\infty_0( M)$ and $f_{\rm m}\in C^\infty( M)$
 such that $f_{\rm i}\big|_{\Lambda_{\rm i}}=1$, 
$f_{\rm m}\big|_{M\setminus\Lambda_{\rm m}} =1$ and
supp$(f_{\rm i} \circ \Phi^{-t}_p) \cap$ supp$(f_{\rm m}\circ\pi_p) =\emptyset$ 
for all $t\in [0,T_m^\delta]$. 
Egorov's Theorem implies that there is  a $C<\infty$ such that for $t\in[0,T_m^\delta]$ 
\[
\left\|  U_1\left(\frac{t}{\epsi}\right) (f_{\rm i}\circ\pi_p)^{\rm W,\epsi}
U_1\left(-\frac{t}{\epsi}\right) - (f_{\rm i} \circ \Phi^{-t}_p)^{\rm W,\epsi} \right\|_{\mathcal{L}
(L^2)}
\leq C\epsi^2\,.
\]
Since, by construction, supp$(f_{\rm i} \circ \Phi^{-t}_p) \cap$ supp$(f_{\rm m}
\circ \pi_p) =\emptyset$,
we have $(f_{\rm m}\circ\pi_p)^{\rm W,\epsi}  (f_{\rm i} \circ \Phi^{-t}_p)^{\rm W,\epsi} = O(\epsi^2)$ by
Corollary \ref{product}. This and the
fact  that $P_{\rm i} = (f_{\rm i}\circ\pi_p)^{\rm W,\epsi} P_{\rm i}$ and that
$(1- P_{\rm m}) =  (1- P_{\rm m}) (f_{\rm m}\circ\pi_p)^{\rm W,\epsi}$ implies (\ref{ipu}).

\hfill $\qed$\medskip

\section{Convergence of the unitary groups}

We prove Theorem \ref{MT}.
Let $\Lambda_{\rm i}+\delta\subset \Lambda_{\rm g}$. 
In the following $T<\infty$ with $T\leq T_{\rm m}^\delta(\Lambda_{\rm i},
\Lambda_{\rm g})$ will be fixed once and for all and we will 
 always assume that $0\leq t \leq T$.

For reasons that will become clear during the proof we have to introduce further sets in
momentum space. Let
\[
\Lambda_{\rm m} = \bigcup_{t\in[0,T], x\in \IR^d}\,
{\rm supp}\left( \1I_{\Lambda_{\rm i}} \circ \Phi_p^{-t}(x,\cdot)\right)
\]
be the subset in momentum space that is reached by the classical dynamics starting
from $\Lambda_{\rm i}$ in our relevant time span. By construction we have that
$\Lambda_{\rm m} +\delta \subseteq \Lambda_{\rm g}$ and we define
$\Lambda_{\rm m1} := \Lambda_{\rm m}+ \delta/ 4$, 
$\Lambda_{\rm m2} := \Lambda_{\rm m} +  \delta/ 2$ and 
$\Lambda_{\rm m3} := \Lambda_{\rm m}+3 \delta/ 4$.
The respective projections  will be denoted by $P_{\rm m1}$, $P_{\rm m2}$ and  
$P_{\rm m3}$. The following Lemma  ensures that
$V^\epsi$ maps wave functions supported in $\Lambda_{\rm m1}$  (resp.\ in $\Lambda_{\rm m2}$,  
$\Lambda_{\rm m3}$) 
to wave functions supported in $\Lambda_{\rm m2}$ (resp.\ in $\Lambda_{\rm m3}$, $\Lambda_{\rm g}$)
up to errors of arbitrary order in $\epsi$. We will use this result in the following implicitly 
many times.

\begin{mathe}{Lemma} \label{VcompL}
Let $\widehat V \in L^1(\IR^d)$ such that $\int dk\, |k|^n |\widehat V(k)|<\infty$
for all $n\in \IN$. Then for each compact $\Lambda\subset \IR^d$ and 
all $m\in \IN$ and $\delta>0$ there is a $C_{\delta,m}$ such that
\[
\left\| \left( V^\epsi - P_{\Lambda+\delta} V^\epsi \right) P_\Lambda
 \right\|_{\mathcal{L}(L^2(M))} \leq \,C_{\delta,m} \epsi^m\,.
\]
\end{mathe}
\noindent {\bf Proof.}
Let $\phi\in$ Ran$P_{\Lambda}$, i.e.\ supp$\phi\subset\Lambda$, and note that
\begin{eqnarray} 
(V^\epsi\phi)(p) &=&
\int dk\, \widehat V(k)\phi(p-\epsi k) \nonumber\\ 
&=& \int\limits_{|k|\leq \epsi^{-1/2}}  dk\, \widehat V(k)\phi(p-\epsi k)
+  \int\limits_{|k|>\epsi^{-1/2}}  dk\, \widehat V(k)\phi(p-\epsi k)\,.\nonumber\\ \label{Vcomp}
\end{eqnarray} 
The first term in (\ref{Vcomp}) is supported in $\Lambda+\delta$ for $\epsi$ sufficiently small.
For the second term note that
\[
 \int\limits_{|k|>\epsi^{-1/2}}  dk\, \widehat V(k)\phi(p-\epsi k) =
 \int\limits_{|k'|>\epsi^{1/2}}  dk'\, \epsi^{-d} \widehat V(k'/\epsi)\phi(p- k')
\]
amounts to convolution with the function $ \1I_{|k'|>\epsi^{1/2}}  \epsi^{-d} \widehat V(k'/\epsi)$ and
therefore, by Young's inequality, the $\mathcal{L}(L^2)$-norm of the corresponding map is bounded by
\[
 \int\limits_{|k'|>\epsi^{1/2}}  dk'\, \epsi^{-d} |\widehat V(k'/\epsi)| =
 \int\limits_{|k|>\epsi^{-1/2}}  dk\,     |k|^{-2m} |k|^{2m}  |\widehat V(k)| \leq \,C_m \epsi^{m}\,.
\]
 
\hfill $\qed$ \medskip

We turn to the proof of (\ref{MTe}).
Using Lemma \ref{Impulserhaltung} together with
$\mathcal{U}\mathcal{U}^*=\,$Id on Ran$P_{\rm m2}$ we get
\begin{eqnarray}
\lefteqn{\left(
U\left(\frac{t}{\epsi}\right) - 
\mathcal{U}^*U_1 \left(\frac{t}{\epsi}
\right)\mathcal{U}\right) P_{\rm i}
 = } \nonumber\\ & = &  
-i U\left(\frac{t}{\epsi}\right) \int_0^{\frac{t}{\epsi}} ds \,
U(-s) \left( H\mathcal{U}^* - \mathcal{U}^* H_1 \right)
U_1(s)\mathcal{U}\, P_{\rm i} \nonumber\\
& = &
-i U\left(\frac{t}{\epsi}\right) \int_0^{\frac{t}{\epsi}} ds\,
U(-s) \left( H - \mathcal{U}^* H_1 \mathcal{U} \right)
P_{\rm m2}\, \mathcal{U}^*U_1(s)\mathcal{U}\,P_{\rm i} + O(\epsi)\,.
\nonumber\\\label{DI1}
\end{eqnarray}
For the last equality note also that a factor of order $O(\epsi^2)$
in the otherwise uniformly bounded integrand leads to the integral being $O(\epsi)$.
We would be done by the same argument, if we could show that
$H- \mathcal{U}^* H_1 \mathcal{U}$ acting on Ran$P_{\rm m2}$ is $O(\epsi^2)$. 
However, the first order term does not vanish, as we will see, and we have to treat
the integral more carefully.

In order to separate the leading order term we write
\[
H- \mathcal{U}^* H_1 \mathcal{U} = \left(H-H_{\rm diag}\right) + 
\left(H_{\rm diag} - \mathcal{U}^* H_1 \mathcal{U}\right)\,,
\]
where
\begin{equation}
H_{\rm diag} = H_0 + P_{\rm g} V^\epsi P_{\rm g} + P^\perp_{\rm g} V^\epsi P^\perp_{\rm g}
\end{equation}
and $P^\perp_{\rm g} := {\bf 1}_\mathcal{H} - P_{\rm g}$.

We will treat the easy part first and show in Lemma \ref{HdiagH1} that the difference 
$H_{\rm diag} - \mathcal{U}^* H_1 \mathcal{U}$
vanishes sufficiently fast on Ran$P_{\rm m2}$. 

However, to do that we need to show that $P_0(p)$ is twice continuously differentiable 
as a function of $p$ and, for
later purposes, we give an explicit expression for $\nabla_pP_0(p)$.

\begin{mathe}{Lemma} \label{gradPLemma}
$P_0(\cdot)\in C^2(\Lambda_{\rm g}, \mathcal{L}(\mathcal{H}_{\rm f}))$ and for $p\in \Lambda_{\rm g}$
one has
\begin{equation} \label{gradP}
\big(\nabla_p P_0\big)(p) = - \big(R_{E}Q_0\big)(p)\big(\nabla_p H_0\big)(p) P_0(p) - 
P_0(p)\big(\nabla_p H_0\big)(p) \big(R_{E}Q_0\big)(p)
\,,
\end{equation}
where  
$\big(R_{E} Q_0\big)(p) = \big(H_0(p) - E(p)\big)^{-1} Q_0(p)$ and 
$Q_0(p) =  {\bf 1}_{ {\mathcal H}_{\rm f} } - P_0(p)$.
\end{mathe}

Note that $(R_{E} Q_0)(p)$ is bounded since $E(p)$ is an isolated eigenvalue for all
$p\in \Lambda_{\rm g}$ and $Q_0(p)$ projects on the orthogonal complement of the corresponding
eigenvector.\smallskip

\noindent {\bf Proof.} \quad 
For $p\in \Lambda_{\rm g}$ we can express
$P_0(p)$ as a contour integral: 
\[
P_0 (p) = -\frac{1}{2\pi i} \oint_{c(p)}d\lambda\, R_\lambda (H_0(p))\,,
\]
where $c(p)$ is a smooth curve in the complex plane circling the isolated eigenvalue $E(p)$ only and
$R_\lambda (H_0(p)) = (H_0(p) - \lambda)^{-1} $.
 Multiplying
\begin{eqnarray*}
0& =& \nabla_p {\bf 1} = \nabla_p (H_0(p) - \lambda)R_\lambda(H_0(p)) \\&=&
(\nabla_pH_0(p))R_\lambda(H_0(p)) + (H_0(p) -\lambda)(\nabla_pR_\lambda(H_0(p)))
\end{eqnarray*}
with $R_\lambda$ from the left we get
\[
\nabla_pR_\lambda(H_0(p)) =- R_\lambda(H_0(p))(\nabla_pH_0(p))R_\lambda(H_0(p))\,,
\]
which is bounded uniformly for $\lambda\in c(p)$ by our assumptions on the family $H_0(p)$.
Hence
\[
\nabla_p P_0 (p) = -\frac{1}{2\pi i} \oint_{c(p)}d\lambda\, \nabla_p R_\lambda (H_0(p))\,.
\]
Collecting the above information we obtain
\begin{eqnarray*}
Q_0(p) \nabla_p P_0 (p) &=& Q_0(p) \nabla_p P_0 (p) (P_0(p)+Q_0(p)) \\
& = & \frac{1}{2\pi i} \oint_{c(p)}d\lambda\, Q_0(p) 
R_\lambda(H_0(p))(\nabla_pH_0(p))R_\lambda(H_0(p)) P_0(p)\,\\
 &=& \frac{1}{2\pi i} \oint_{c(p)}  d\lambda\,
R_\lambda(H_0(p))Q_0(p)(\nabla_pH_0(p))\frac{1}{E(p) -\lambda} P_0(p)\,\\
&=& - R_{E}(H_0(p))Q_0(p) \left(\nabla_p H_0(p)\right)P_0(p)\,,
\end{eqnarray*}
where the $Q_0(p) \nabla_p P_0 (p) Q_0(p)$ vanishes, since in this case the integrand
is an analytic function inside of $c(p)$.
Therefore $Q_0(p) \nabla_p P_0 (p) = Q_0(p) \nabla_p P_0 (p)P_0(p)$ and, using self adjointness
of $\nabla_p P_0(p)$, it follows that  
\[
P_0(p) \nabla_p P_0 (p) = P_0(p) \nabla_p P_0 (p)Q_0(p) =  
\left( - R_{E}(H_0(p))Q_0(p) \left(\nabla_p H_0(p)\right)P_0(p)\right)^*.
\]
This yields (\ref{gradP}) and 
$P_0(\cdot)\in C^2(\Lambda_{\rm g}, \mathcal{L}(\mathcal{H}_{\rm f}))$
by applying the same argument again to  (\ref{gradP}).

\hfill $\qed$\medskip

\begin{mathe}{Lemma} \label{HdiagH1}
The phase of  $\psi_0(p)$, $p\in\Lambda_{\rm g}$, can be chosen such that
$\psi_0(\cdot)\in C^2(\Lambda_{\rm g},\mathcal{H}_{\rm f})$. 
Then for $\epsi$ sufficiently small there is a $C<\infty$ such that
\begin{equation}
\Big\| \big( H_{\rm diag} -  \mathcal{U}^* H_1 \mathcal{U}\big) P_{\rm m2}
\Big\| \leq C\epsi^2\,.
\end{equation}
\end{mathe}
\noindent {\bf Proof.} \quad Note that Lemma \ref{VcompL} implies that 
\[
\| P_{\rm g}^\perp H_{\rm diag} P_{\rm m2} \|  \leq \,C\epsi^2\,, 
\] 
and we have by construction that 
$ P_{\rm g}^\perp \mathcal{U}^* H_1 \mathcal{U}  P_{\rm m2} =0$. Hence it suffices to consider
the difference projected onto Ran$P_{\rm g}$.
On Ran$P_{\rm m2}$ we have
\begin{eqnarray*} \lefteqn{ P_{\rm g}
(H_{\rm diag} \phi\psi_0)(p)\, \, =}
\\ &&  E(p)\phi(p)\psi_0(p)+\1I_{\Lambda_{\rm g}}(p)\, \int dk\,
\widehat V (k) \phi(p-\epsi k)\,
\langle \psi_0(p),\psi_0(p-\epsi k)\rangle_{\mathcal{H}_{\rm f}}\, \psi_0(p)
\end{eqnarray*}
and 
\[
(\mathcal{U}^*H_1\mathcal{U} \phi\psi_0)(p) =  E(p)\phi(p)\psi_0(p)+\1I_{\Lambda_{\rm g}}(p)\,
 \int dk\, \widehat V (k) \phi(p-\epsi k)
\,\psi_0(p)\,.
\]
Hence 
\begin{eqnarray} \lefteqn{\hspace{-.1cm} P_{\rm g}
\left( H_{\rm diag} - \mathcal{U}^*H_1\mathcal{U}\right)( \phi\psi_0)(p) =}
\\
&=&\1I_{\Lambda_{\rm g}}(p)\,
\int dk\, \widehat V (k) \phi(p-\epsi k)\,\left(
\langle \psi_0(p),\psi_0(p-\epsi k)\rangle_{\mathcal{H}_{\rm f}}-1\right)\, \psi_0(p)\,.
\nonumber
\end{eqnarray}
We will show  that there is a constant $C$ such that
\begin{equation} \label{HdHe}
\left| \langle \psi_0(p),\psi_0(p-\epsi k)\rangle_{\mathcal{H}_{\rm f}}-1\right| \leq C |k|^2\epsi^2 
\end{equation}
for all $p\in \Lambda_{\rm g}$ and $k$ with $p-\epsi k \in \Lambda_{\rm g}$. Therefore
\begin{eqnarray*}
\Big\| 
\left( H_{\rm diag} - \mathcal{U}^*H_1\mathcal{U}\right) \phi\psi_0
\Big\|_\mathcal{H} &\leq& 
C\epsi^2\int dk\, | 
\widehat V(k)|\,|k|^2 
\big\|\phi(\cdot-\epsi k)\psi_0(\cdot)\big\|_\mathcal{H} \,
\\
&\leq& C \epsi^2 
\|\phi\|_{L^2(M)}\,
\int dk\, |\widehat V(k)|\,|k|^2\,
\\
& = &  
C \epsi^2 \|\phi\psi_0\|_\mathcal{H}\,\int dk\, |\widehat V(k)|\,|k|^2\,   .
\end{eqnarray*} 

To see (\ref{HdHe}), note that for 
$\psi_0(\cdot)\in C^2(\Lambda_{\rm g},\mathcal{H}_{\rm f})$
 Taylor expansion yields
\[
\psi_0(p-\epsi k) = \psi_0(p) -\epsi  k\cdot\nabla_p\psi_0(p) + \frac{1}{2}\epsi^2
  k\cdot H(\psi_0)(p')k\,,
\]
where $H(\psi_0)$ denotes the Hessian and  $\frac{1}{2}\epsi^2  k\cdot H(\psi_0)(p')k$ is the
Lagrangian remainder. In view of 
$\langle \psi_0(p),\nabla_p\psi_0(p)\rangle_{\mathcal{H}_{\rm f}}=0$, which follows from 
comparing (\ref{gradP}) with
\[
(\nabla_pP_0 \psi)(p) = \langle \psi_0(p), \psi \rangle_{\mathcal{H}_{\rm f}}
\nabla_p\psi_0(p) + \langle \nabla_p\psi_0(p), \psi \rangle_{\mathcal{H}_{\rm f}}
\psi_0(p)\,,
\]
we obtain 
\[ 
\left| \langle \psi_0(p),\psi_0(p-\epsi k)\rangle_{\mathcal{H}_{\rm f}}-1\right| 
\leq C(p) |k|^2\epsi^2 \,.
\]
Here $C(p) = \frac{1}{2} \sum_{i,j} | \langle \psi_0(p'), \partial_{p_i} \partial_{p_j}
\psi_0(p')\rangle |$, which is bounded uniformly for $p\in\Lambda_{\rm g}$ by continuity.

We are left to show that one can indeed choose the phase of the eigenfunctions
such that $\psi_0(\cdot)\in C^2(\Lambda_{\rm g},\mathcal{H}_{\rm f})$.
Since $P_0(\cdot)\in C^2(\Lambda_{\rm g},\mathcal{L}(\mathcal{H}_{\rm f}))$, 
one can  cover $\Lambda_{\rm g}$ with finitely many open sets $U_n$, each  
containing a  $p_n$ such that  
$\psi^n_0(p) = P_0(p)\psi_0(p_n)/ \| P_0(p)\psi_0(p_n) \|$ is well defined for all 
$p\in\overline U_n$. It is now straightforward to connect these pieces in order
to get a $C^2$ version of $\psi_0(p)$ on all of $\Lambda_{\rm g}$. 
Note that if we would replace smoothness by analyticity, this last step
would become nontrivial, in particular, if $M$ is a torus
(cf., e.g., \cite{Nenciu}).

\hfill $\qed$\medskip

With the help of Lemma \ref{HdiagH1} and (\ref{DI1}), we have
\begin{eqnarray*}
\lefteqn{\left(
U\left(\frac{t}{\epsi}\right) - 
\mathcal{U}^*U_1 \left(\frac{t}{\epsi}
\right)\mathcal{U}\right) P_{\rm i}
 = } \\
& = &
-i U\left(\frac{t}{\epsi}\right) \int_0^{\frac{t}{\epsi}} ds\,
U(-s) \left( H - H_{\rm diag} \right)
P_{\rm m2}\, \mathcal{U}^*U_1(s)\mathcal{U}\,P_{\rm i} + O(\epsi)\,.
\end{eqnarray*}
In the following lemma we isolate the leading order term
in $(H-H_{\rm diag})P_{\rm m2}$.

\begin{mathe}{Lemma} \label{VODLemma}
\begin{equation} \label{VOD}
(H-H_{\rm diag})P_{\rm m2} = -\epsi Q_{\rm g}
(\nabla_p P_{\rm g}) P_{\rm m3} \cdot F^\epsi P_{\rm m2} + O(\epsi^2)\,,
\end{equation}
where
 $(\nabla_p P_{\rm g}) = \int^\oplus_{\Lambda_{\rm g}}dp\, \nabla_p P_0(p)$, 
$Q_{\rm g} = \int^\oplus_{\Lambda_{\rm g}}dp\,Q_0(p)$ and
\[
(F^\epsi \psi)(p)  := - i \int  dk\, k\,\widehat{V}(k)
\psi(p-\epsi k) \,.
\]  
\end{mathe}

\noindent{\bf Proof.} Using Lemma \ref{VcompL} we obtain
\begin{eqnarray*}\lefteqn{
(H-H_{\rm diag}) P_{\rm m2} \psi   =  P_{\rm g}^\perp V^\epsi P_{\rm m2} \psi}
\\
& =&  \int^\oplus_{\Lambda_{\rm m3}}  dp\,\left( Q_0(p)
\int  dk\,\widehat V (k) P_0(p-\epsi k) (P_{\rm m2} \psi)(p-\epsi k)\, \right)+O(\epsi^2)\,.
\end{eqnarray*}
Since $P_0: \Lambda_{\rm g} \to \mathcal{L}(\mathcal{H}_{\rm f})$ is twice continuously differentiable,
we have that 
\begin{equation}
P_0(p-\epsi k) = P_0(p) -\epsi k\cdot (\nabla_p P_0)(p) + \epsi^2\, k\cdot
H(P_0)(p'(p,\epsi k))\cdot k\,,
\end{equation}
where the last term is the Lagrangian remainder with $H$ denoting the Hessian. Hence,
for $p\in\Lambda_{\rm g}$,
\begin{eqnarray}
\lefteqn{ 
\int  dk\,\widehat V (k) P_0(p-\epsi k) (P_{\rm m2} \psi)(p-\epsi k)\,} \nonumber\\
& = & \int  dk \,\widehat V (k) \big( P_0 (p)-\epsi k\cdot (\nabla_p P_0)(p)\big) 
(P_{\rm m2} \psi)(p-\epsi k)\,\label{h1} \\
&  & +\,\, \epsi^2 \int  dk\, \widehat V (k)\, k\cdot
H(P_0)(p'(p,\epsi k))\cdot k \,(P_{\rm m2} \psi)(p-\epsi k)\,. \label{h2}
\end{eqnarray}
Since
\begin{eqnarray} \label{est} \lefteqn{
\left\|\1I_{\Lambda_{\rm m3}}(\cdot)\, \int  dk\, \widehat V (k)\, k\cdot
H(P_0)(p'(\cdot,\epsi k))\cdot k \,(P_{\rm m2} \psi)(\cdot-\epsi k)\,\right\|_\mathcal{H} }
\\
&\leq&
 \int dk\, \widehat V (k)\, \big\|\1I_{\Lambda_{\rm m3}}(\cdot)\, k\cdot
H(P_0)(p'(\cdot,\epsi k))\cdot k \,(P_{\rm m2} \psi)(\cdot-\epsi k)\,\big\|_\mathcal{H}
\\
&\leq & \sup_{p\in \Lambda_{\rm m2}}\| H(P_0)(p)\| \int dk\, |\widehat V(k)|\, |k|^2 \,
\|\1I_{\Lambda_{\rm m3}}(\cdot)\, (P_{\rm m2} \psi)(\cdot -\epsi k) \|_{\mathcal{H}}\, \nonumber\\
&\leq&
C\,\|\psi\|_{\mathcal{H}}\,\int dk\,|\widehat V(k)|k^2\,,\nonumber
\end{eqnarray}
(\ref{h2}) is $O(\epsi^2)$ in $\mathcal{L}(\mathcal{H}_{\rm f})$
 and multiplying (\ref{h1}) with $ Q_0(p)$ from the left
establishes (\ref{VOD}).

\hfill$\qed$\medskip

Including Lemma \ref{VODLemma} we are left with
\begin{eqnarray*}
\lefteqn{\left(
U\left(\frac{t}{\epsi}\right) - 
\mathcal{U}^*U_1 \left(\frac{t}{\epsi}
\right)\mathcal{U}\right) P_{\rm i}
 = } \\
& = &
 i \epsi U\left(\frac{t}{\epsi}\right) \int_0^{\frac{t}{\epsi}} ds\,
U(-s) Q_{\rm g} (\nabla_p P_{\rm g})P_{\rm m3}\cdot F^\epsi 
P_{\rm m2}\, \mathcal{U}^*U_1(s)\mathcal{U}\,P_{\rm i} + O(\epsi)\,.
\end{eqnarray*}

To exploit the time averaging 
 we write $ Q_{\rm g} (\nabla_p P_{\rm g})$ as a time derivative, 
at least in approximation.
Let 
\[
B(p) := R_{E(p)}(H_0(p))^2 Q_0(p) (\nabla_pH_0)(p)P_0(p)\,.
\]

\begin{mathe}{Lemma} \label{BH}
\begin{eqnarray}
Q_{\rm g} (\nabla_p P_{\rm g})P_{\rm m3}& = &[B, H_0] P_{\rm m3} \label{B1} \\
&=& [B,H] P_{\rm m3} + O(\epsi) \label{B2}
\end{eqnarray}
 and
\begin{equation}\label{FH}
[F^\epsi, H]P_{\rm m2} = O(\epsi)
\end{equation}
in $\mathcal{L}(\mathcal{H})$.
\end{mathe}

\noindent {\bf Proof.}\quad 
Recalling (\ref{gradP}), we have for $p\in \Lambda_{\rm g}$ that
\[
Q_0(p) (\nabla_p P_0)(p)P_0(p) = 
-  R_{E(p)}(H_0(p)) Q_0(p) (\nabla_pH_0)(p)P_0(p)
\]
and thus (\ref{B1}) follows from direct computation,
\begin{eqnarray*}
\lefteqn{
B(p)H_0(p) - H_0(p)B(p) =}\\
&=& 
-(H_0(p) - E(p)) R_{E(p)}(H_0(p))^2 Q_0(p) (\nabla_pH_0)(p)P_0(p)\\
&=&
-  R_{E(p)}(H_0(p)) Q_0(p) (\nabla_pH_0)(p)P_0(p)\\
&=&
Q_0(p) (\nabla_p P_0)(p)P_0(p)\,.
\end{eqnarray*}
For (\ref{B2}) we need to show that $[B,V^{\epsi}]P_{\rm m3}= O(\epsi)$:
\[
[B, V^\epsi]P_{\rm m3} =
[ R_{E}(H_0)^2 Q_{\rm g} (\nabla_pH_0)P_{\rm g}, V^\epsi]P_{\rm m3}\,.
\]
It follows from the proof of Lemma \ref{VODLemma} and using Lemma \ref{VcompL} that
\begin{eqnarray*}
[P_{\rm g},V^\epsi]P_{\rm m3}& =& P_{\rm g} V^\epsi  P_{\rm m3} - P_{\rm g}  V^\epsi P_{\rm g} 
P_{\rm m3}
-  Q_{\rm g}  V^\epsi P_{\rm g} P_{\rm m3}+O(\epsi) \\&=& - Q_{\rm g}  V^\epsi P_{\rm g} P_{\rm m3}
+O(\epsi)= O(\epsi)\,.
\end{eqnarray*}
Hence
\[
[B, V^\epsi]P_{\rm m3} =
[ R_{E}(H_0)^2 Q_{\rm g} (\nabla_pH_0), V^\epsi] P_{\rm g} P_{\rm m3} + O(\epsi)\,.
\]
Using again Lemma \ref{VcompL} and the fact that also $\nabla_pH_0$ acts on
$p$-fibers, we observe that
\begin{eqnarray} \label{H0comm} \lefteqn{
\big([\nabla_pH_0, V^\epsi] P_{\rm m3} \psi\big)(p) = \1I_{\Lambda_{\rm g}}(p)
\big([\nabla_pH_0, V^\epsi] P_{\rm m3} \psi\big)(p) + O(\epsi)} \nonumber\\
& =&\1I_{\Lambda_{\rm g}}(p)
\int dk\,
\widehat V (k) \big( ( \nabla_pH_0)(p) -  ( \nabla_pH_0)(p-\epsi k)  \big) 
(P_{\rm m3} \psi)(p-\epsi k) + O(\epsi)\nonumber\\
& = & \epsi\1I_{\Lambda_{\rm g}}(p)\int dk\,\widehat V (k) \,k\cdot H(H_0)(p'(k))
(P_{\rm m3} \psi)(p-\epsi k) + O(\epsi)
\end{eqnarray}
where $H(H_0)(p'(k))$ denotes again the Hessian evaluated  at the appropriate point $p'(k)$.
 By assumption we have that $H(H_0)(p)$ is bounded  uniformly for $p\in \Lambda_{\rm g}$
in the graph norm, which  is equivalent to the $\mathcal{H}$-norm on
span$\{ \psi_0(p), p\in \Lambda_{\rm g}\}$.
Thus the above expression is $O(\epsi)$ as $\epsi \to 0$ (cf.\ estimate (\ref{est})).
Now
\[
[B, V^\epsi]P_{\rm m3} =
[ R_{E}(H_0)^2 Q_{\rm g}, V^\epsi] (\nabla_pH_0) P_{\rm m3} + O(\epsi)\,.
\]
Since $ R_{E(p)}(H_0(p))^2 Q_0(p)$ is bounded and differentiable with respect to $p$,
one can show by the same line of arguments as in the case of $[P_{\rm g},V^\epsi]P_{\rm m2}$ 
(cf.\ proof of Lemma \ref{VODLemma}) that
\[
[ R_{E}(H_0)^2 Q_{\rm g},V^\epsi] (\nabla_pH_0) P_{\rm m3}      = O(\epsi)\,.
\]
Thus we showed (\ref{B2}).
For (\ref{FH}) first observe that $[F^\epsi,V^\epsi]=0$.
$[F^\epsi, H_0] = O(\epsi)$ follows from an estimate analogous to (\ref{H0comm}).

\hfill $\qed$\medskip

Let $A = B\cdot P_{\rm m3} F^\epsi$. Then Lemma \ref{BH} yields
\[
Q_{\rm g}(\nabla_p P_{\rm g})\cdot P_{\rm m3} F^\epsi P_{\rm m2} =
[A,H]P_{\rm m2} + O(\epsi)
\]
in $\mathcal{L}(\mathcal{H})$, where $[P_{\rm m3} F^\epsi,H]P_{\rm 2} = O(\epsi)$
follows immediately from (\ref{FH})
Let $A(t)= U(-t) A U(t)$, then
\begin{eqnarray*}
\lefteqn{\left(
U\left(\frac{t}{\epsi}\right) - 
\mathcal{U}^*U_1 \left(\frac{t}{\epsi}
\right)\mathcal{U}\right)\, P_{\rm i}
 = } \\
& = &
i \epsi U\left(\frac{t}{\epsi}\right) \int_0^{\frac{t}{\epsi}} ds\,
U(-s) [A,H] U(s)U(-s)
P_{\rm m2}\, \mathcal{U}^*U_1(s)\mathcal{U}\, P_{\rm i} + O(\epsi) \, \\
&=&
 - \epsi U\left(\frac{t}{\epsi}\right) \int_0^{\frac{t}{\epsi}} ds\,
\left( \frac{d}{ds} A(s)\right) U(-s)
P_{\rm m2}\, \mathcal{U}^*U_1(s)\mathcal{U}\, P_{\rm i} + O(\epsi) \, \\
&=&
- \epsi U\left(\frac{t}{\epsi}\right)\Big[
A(s) U(-s)
P_{\rm m2}\, \mathcal{U}^*U_1(s)\mathcal{U}\Big]^\frac{t}{\epsi}_0\, 
P_{\rm i} + O(\epsi)  \\
&&+\,  \epsi U\left(\frac{t}{\epsi}\right) \int_0^{\frac{t}{\epsi}} ds\,
 A(s)\frac{d}{ds}\Big( U(-s)
P_{\rm m2}\, \mathcal{U}^*U_1(s)\mathcal{U}\Big)\, P_{\rm i}  \\
&=&  i \epsi U\left(\frac{t}{\epsi}\right) \int_0^{\frac{t}{\epsi}} ds\,
 A(s) U(-s) \Big(H P_{\rm m2}\mathcal{U}^*  -P_{\rm m2}\mathcal{U}^* H_1 \Big)
\, U_1(s)\mathcal{U}\, P_{\rm i} + O(\epsi) \\
&=&  i \epsi U\left(\frac{t}{\epsi}\right) \int_0^{\frac{t}{\epsi}} ds\,
 A(s) U(-s) \Big(H   - \mathcal{U}^* H_1\mathcal{U} \Big)P_{\rm m1}
\, \mathcal{U}^*U_1(s)\mathcal{U}\, P_{\rm i} + O(\epsi)\,.
\end{eqnarray*}
For the last equality we used again Lemma \ref{Impulserhaltung} and the fact that
Lemma \ref{VcompL} guarantees 
$P_{\rm m2}  \mathcal{U}^* H_1\mathcal{U} P_{\rm m1} =  
\mathcal{U}^* H_1\mathcal{U} P_{\rm m1}+O(\epsi^2)$.
Finally  also the last integral is $ O(\epsi)$, since we showed already that 
$(H   - \mathcal{U}^* H_1\mathcal{U})P_{\rm m1}= O(\epsi)$ and the proof of
Theorem \ref{MT} is completed.

\section{Convergence of the macroscopic observables}

In this section we prove Theorem \ref{effop} and Theorem \ref{effop2}. We start by 
showing 

\begin{mathe}{Lemma} \label{gcor}
Let $g\in C^\infty(M)$ such that 
$g\circ\pi_p\in S^0_0(1)$ and let the assumptions of Theorem \ref{effop}
be satisfied. Then
\begin{equation}\label{peff} 
\Big\| \Big( U(-t/\epsi) \left( g(p)\otimes {\bf 1} \right)  U(t/\epsi) -  \mathcal{U}^*
 U_1(-t/\epsi) g(p) U_1(t/\epsi)\mathcal{U} \Big) P_{\rm i}\Big\|
\leq \,C\,\epsi\,, 
\end{equation}
where $g(p):= (g\circ\pi_p)^{\rm W,\epsi}$ denotes the operator of multiplication 
with the function $g$ on $L^2(M)$. 
\end{mathe}

\noindent {\bf Proof}.\quad
 Let $\Lambda_{\rm i,\gamma} = \Lambda_{\rm i}+\gamma$
and $P_{\rm i,\gamma}$ the corresponding projection.
Let $\gamma$ be sufficiently small to ensure that 
supp$\big((\1I_{\Lambda_{\rm i},\gamma}\circ\Phi^{-t}_p)(x,\cdot)\big)
+\delta/2\subset \Lambda_{\rm g}$ for
all $t\in [0,T^\delta_{\rm m}]$ and $x\in \IR^d$.
We abbreviate $ g(t) := U(-t/\epsi) (g(p)\otimes{\bf 1}) U(t/\epsi)$ and $g_1(t):= U_1(-t/\epsi) g(p) U_1(t/\epsi)$
and
split (\ref{peff}) into two parts:
\begin{eqnarray}
\Big\| \Big(g(t)-\mathcal{U}^*g_1(t)\mathcal{U}\Big) 
P_{\rm i} \Big\| & \leq &
\Big\|P_{\rm i,\gamma} \Big(g(t)-\mathcal{U}^*g_1(t)\mathcal{U}\Big) 
P_{\rm i} \Big\|\label{peff1} \\
&& +\,
\Big\|P_{\rm i,\gamma}^\perp \Big(g(t)-\mathcal{U}^*g_1(t)\mathcal{U}\Big) 
P_{\rm i} \Big\|\,.\label{peff2}
\end{eqnarray}
Using $\mathcal{U}\mathcal{U}^*P_{\rm g}=P_{\rm g}$ on $L^2(M)$ 
we find  that (\ref{peff1}) can be estimated
as 
\begin{eqnarray*}\lefteqn{
\Big\|P_{\rm i,\gamma} \Big(g(t)-\mathcal{U}^* g_1(t)\mathcal{U}\Big) 
P_{\rm i} \Big\|}\\ &\hspace{-.1cm}\stackrel{\rm Lemma\, \ref{Impulserhaltung}}{\leq}&
\hspace{-.45cm}\Big\|P_{\rm i,\gamma} \Big(
 U\left(-t/\epsi\right)\mathcal{U}^*-\mathcal{U}^*
 U_1\left(-t/\epsi\right)
\Big)\mathcal{U}\mathcal{U}^*P_{\rm g} \,g(p)\, U_1\left(t/\epsi\right)\mathcal{U}
P_{\rm i} \Big\|+O(\epsi^2)\\
&&+\,
\Big\|P_{\rm i,\gamma} U\left(-t/\epsi\right)\,(g(p)\otimes{\bf 1})\, \Big(
 U\left(t/\epsi\right)-\mathcal{U}^*
 U_1\left(t/\epsi\right)\mathcal{U}
\Big)
P_{\rm i}\Big\|\\
&\hspace{-.1cm}\leq &\hspace{-.45cm}
\Big\|P_{\rm i,\gamma} \Big(
 U\left(-t/\epsi\right)-\mathcal{U}^*
 U_1\left(-t/\epsi\right)\mathcal{U}
\Big)\Big\| \,\Big\| P_{\rm g}\,g(p)\, U_1\left(t/\epsi\right)\mathcal{U}
P_{\rm i} \Big\|+O(\epsi^2)\\
&&+\,
\Big\|P_{\rm i,\gamma} U\left(-t/\epsi\right)\,(g(p)\otimes {\bf 1})\,P_{\rm g} \Big\|\,\Big\| \Big(
 U\left(t/\epsi\right)-\mathcal{U}^*
 U_1\left(t/\epsi\right)\mathcal{U}
\Big)P_{\rm i} \Big\|\\
&\stackrel{\rm Theorem\,\ref{MT}}{=}& \,O(\epsi)\,.
\end{eqnarray*}
It remains to show that in (\ref{peff2}) both terms are of order $O(\epsi)$ separately.
By construction we can pick $f_{\rm i}$, $f^\perp_{\rm i,\gamma} \in C^\infty(M)$
such that $f_{\rm i}\big|_{\Lambda_{\rm i}} = 1$, 
$f^\perp_{\rm i,\gamma}\big|_{{M\setminus\Lambda_{\rm i,\gamma}}}=1$ and
supp$f_{\rm i}\bigcap$ supp$f^\perp_{\rm i,\gamma} =\emptyset$. 
Using Egorov's Theorem  and Corollary \ref{product} (i)
we get 
\begin{eqnarray*}
g_1(t)\,(f_{\rm i}\circ\pi_p)^{\rm W,\epsi}& =& (g\circ\Phi^t_p)^{\rm W,\epsi}\, 
(f_{\rm i}\circ\pi_p)^{\rm W,\epsi} +O(\epsi^2)\\ & = & 
((g\circ\Phi^t_p)\, (f_{\rm i}\circ\pi_p))^{\rm W,\epsi} +O(\epsi)
\end{eqnarray*}
and therefore, with Corollary \ref{product} (ii), that
\begin{eqnarray*}
\big\| P_{\rm i,\gamma}^\perp \,\mathcal{U}^*\, g_1(t)\,\mathcal{U}\, P_{\rm i}\big\| &\leq&
\big\| ( f^\perp_{\rm i,\gamma}\circ\pi_p)^{\rm W,\epsi}\, g_1(t)\, (f_{\rm i}\circ
 \pi_p)^{\rm W,\epsi}\, \mathcal{U}\,
P_{\rm i}\big\| \\
& = &
\big\| (f^\perp_{\rm i,\gamma}\circ\pi_p)^{\rm W,\epsi} \,
\left((g\circ \Phi^t_p)\, (f_{\rm i}\circ\pi_p)\right)^{\rm W,\epsi}\,\mathcal{U} \,
P_{\rm i}\big\| + O(\epsi)\\
& = & O(\epsi)\,,
\end{eqnarray*}
since $f^\perp_{\rm i,\gamma}\circ\pi_p$ and $(g\circ\Phi^t_p)\,(f_{\rm i}\circ\pi_p)$ 
are disjointly supported.
Finally we compute that
\begin{eqnarray*} \lefteqn{
\big\| P_{\rm i,\gamma}^\perp  g(t)\, P_{\rm i}\big\| = \big\|  P_{\rm i,\gamma}^\perp
 U\left(-t/\epsi\right)\,(g(p)\otimes{\bf 1})\, U\left(t/\epsi\right)
P_{\rm i}\big\|}\\
&\stackrel{\rm Theorem\,\ref{MT}}{=}&
\big\|  P_{\rm i,\gamma}^\perp
 U\left(-t/\epsi\right)\,\mathcal{U}^*\,g(p) U_1\left(t/\epsi\right)
\mathcal{U}
P_{\rm i}\big\|+O(\epsi)\\
&\stackrel{\rm Egorov}{=}&
\big\|  P_{\rm i,\gamma}^\perp
 U\left(-t/\epsi\right)\mathcal{U}^* U_1\left(t/\epsi\right)
 (g\circ\Phi^t_{\rm p})^{\rm W,\epsi} (f_{\rm i}\circ\pi_p)^{\rm W,\epsi}
\mathcal{U}P_{\rm i}\big\|+O(\epsi)\\
&\stackrel{\rm Corollary\,\ref{product}\,(i)}{=}&
\big\|  P_{\rm i,\gamma}^\perp
 U\left(-t/\epsi\right)\mathcal{U}^* U_1\left(t/\epsi\right)
 (f_{\rm i}\circ\pi_p)^{\rm W,\epsi} (g\circ\Phi^t_{\rm p})^{\rm W,\epsi}\mathcal{U}
P_{\rm i}\big\|+O(\epsi)\\
&\leq&
\big\|  P_{\rm i,\gamma}^\perp
 U\left(-t/\epsi\right)\mathcal{U}^* U_1\left(t/\epsi\right)\mathcal{U}
 P_{\rm i,\gamma}\,\mathcal{U}^* (g\circ\Phi^t_{\rm p})^{\rm W,\epsi}
\mathcal{U}P_{\rm i}\big\|+O(\epsi)\\
&\stackrel{\rm Theorem\,\ref{MT}}{=}&
\big\|  P_{\rm i,\gamma}^\perp
 U\left(-t/\epsi\right) U\left(t/\epsi\right)
 P_{\rm i,\gamma} (g\circ\Phi^t_{\rm p})^{\rm W,\epsi}\mathcal{U}P_{\rm i}\big\|+O(\epsi)\\
&=& O(\epsi)\,.
\end{eqnarray*}
For the third equality we inserted $U_1(t/\epsi)U_1(-t/\epsi) ={\bf 1}$ 
in order to apply Egorov's Theorem on $g_1(t)$.

\hfill $\qed$\medskip

We now come to the proof of (\ref{xeff}). We have (at the moment only formally) that
\[
x^\epsi(t) = i\,\epsi  \nabla_p\otimes {\bf 1} \,+\,
\int_0^t ds \, U(-s/\epsi)\, [i\nabla_p\otimes{\bf 1} , H] U(s/\epsi)\,.
\]
On $H^1(\IR^d)\otimes \mathcal{H}_{\rm f}$ this gives
\begin{eqnarray}
x^\epsi(t)P_{\rm i}  &=& \left(i\, \epsi  \nabla_p\otimes{\bf 1}\right)P_{\rm i} +
\int_0^t ds \, U(-s/\epsi)\, [i\nabla_p\otimes{\bf 1} , H] U(s/\epsi) \,P_{\rm i}\,\nonumber\\
&=& \left(i\,
 \epsi  \nabla_p\otimes{\bf 1}\right)P_{\rm i} +\int_0^t ds\, U(-s/\epsi)\,( \nabla_p H_0)
\,P_{\rm g} U(s/\epsi) \,P_{\rm i}\, \nonumber\\
&& \qquad \qquad +\,\int_0^t ds\, U(-s/\epsi)\,( \nabla_p H_0)
\,P_{\rm g}^\perp U(s/\epsi) \,P_{\rm i}\, \nonumber\\
& = &\left( i\, \epsi  \nabla_p\otimes{\bf 1}\right)P_{\rm i} +\int_0^t ds\, U(-s/\epsi)\,\left( i\nabla E(p)\otimes{\bf 1}\right)P_{\rm g}
 U(s/\epsi) \,P_{\rm i}\,\nonumber\\ && \qquad\qquad +\, O(\epsi)\,,\label{Pperp}
\end{eqnarray}
where we prove in a moment that this is well defined on  
$H^1(\IR^d)\otimes \mathcal{H}_{\rm f}$ 
and that the last equality in (\ref{Pperp}) holds. 

On the other hand, again on $H^1(\IR^d)\otimes \mathcal{H}_{\rm f}$,
\[
 \mathcal{U}^*\,x_1^\epsi(t)\, \mathcal{U}\,P_{\rm i} = \mathcal{U}^*\, i\, \epsi  
\nabla_p\,\mathcal{U}\,P_{\rm i}\, +\,
 \int_0^t ds\, \mathcal{U}^*\, U_1(-s/\epsi)\,i\nabla E(p) \, U_1(s/\epsi)\, \mathcal{U}\, 
\,P_{\rm i}\,.
\]
Since $\nabla E\in C_0^\infty(M)$ we can apply Lemma \ref{gcor} and finish by showing that
\begin{equation} \label{heins}
\Big\| \Big(i\, \epsi  \nabla_p\otimes{\bf 1} -  \mathcal{U}^*\, i\, \epsi  \nabla_p\,\mathcal{U}
\Big) P_{\rm i}\Big\| = O(\epsi)\,.
\end{equation}
Let $\psi = \phi(p)\psi_0(p)$, then
\begin{eqnarray*}
\Big\| \Big(i\, \epsi  \nabla_p\otimes{\bf 1} -  \mathcal{U}^*\, i\, \epsi  \nabla_p\,\mathcal{U}
\Big)  \phi\psi_0\Big\|_\mathcal{H}& =& \epsi \, \Big\| \phi \nabla_p \psi_0\Big\|_\mathcal{H} \\
&\leq& \epsi\,C\| \phi\|_{L^2(M)} =   \epsi\,C\| \psi\|_\mathcal{H}\,,
\end{eqnarray*}
which proves (\ref{heins}). To complete the proof of Theorem \ref{effop} we are left to show the
last equality of (\ref{Pperp}), i.e.\ that
\[
 (\nabla_p H_0)
\,P_{\rm g}^\perp U(s/\epsi) \,P_{\rm i} = O(\epsi)
\]
uniformly for $s\in [0,T]$. If $(\nabla_pH_0)$ was a bounded operator, Lemma \ref{Impulserhaltung}
translated to the full dynamics via Theorem \ref{MT}
would take care of that. Recall, however, that by assumption $(\nabla_p H_0) (H_0-i)^{-1}$ is bounded
and hence it suffices to show 
\[
H_0
\,P_{\rm g}^\perp U(s/\epsi) \,P_{\rm i} = O(\epsi)\,.
\]
We calculate
\begin{eqnarray}
H_0\,P_{\rm g}^\perp U(s/\epsi) \,P_{\rm i} &=&
P_{\rm g}^\perp\, H_0\, U(s/\epsi) \,P_{\rm i} \nonumber\\
&\stackrel{(*)}{=}& P_{\rm g}^\perp\, H\, U(s/\epsi) \,P_{\rm i} + O(\epsi) \nonumber\\
&=&  P_{\rm g}^\perp\,  U(s/\epsi)\, H \,P_{\rm i} + O(\epsi) \nonumber\\
&\stackrel{\rm Lemma \,\ref{VcompL}}{=}&  P_{\rm g}^\perp\,  U(s/\epsi)\, P_{\rm i,\gamma} H \,P_{\rm i} + O(\epsi) \nonumber\\
&\stackrel{\rm Lemma\,\ref{Impulserhaltung}}{=}& O(\epsi)\,,\label{xx}
\end{eqnarray}
where (*) follows from Lemma \ref{Impulserhaltung} together with 
$[P_{\rm g}^\perp, V^\epsi]P_{\rm m} = O(\epsi)$ (cf.\ proof of Lemma \ref{BH}) and 
$P_{\rm i,\gamma}$ is defined as in the proof of Lemma \ref{gcor}.

Finally we sketch the proof of Theorem \ref{effop2}. First we conclude from 
Theorem \ref{effop} that in the special case of $a=f\circ \pi_x\in S^0_0(1)\cap 
C_\infty(\IR^d\times M)$, 
for some $f\in C^\infty(\IR^d)$, (\ref{xeff2}) holds.
To see this, note that by the functional calculus for self-adjoint operators it
suffices to show 
\begin{equation}\label{xeff3}
\left\| \left(
f(x^\epsi(t)) - f(\mathcal{U}^*\,x_1^\epsi(t)\,\mathcal{U})
\right) P_{\rm i} \right\| \leq C\,\epsi\,.
\end{equation}
However, (\ref{xeff3}) follows from a standard approximation argument using 
the general Stone-Weierstra\ss\ Theorem and norm-convergence of the resolvents
(cf., e.g., Theorem VIII.20 in \cite{RS1}).

Next we observe that for $a(x,p) = f(x)g(p)$ Lemma \ref{Impulserhaltung}
and Corollary \ref{product} imply (\ref{xeff2}) in a rather straightforward way.

Finally one can approximate general $a\in S^0_0(1)\cap C_\infty(\IR^d\times M)$ 
by products referring again to the
general Stone-Weierstra\ss\ Theorem.

\section{Semi-classical distributions}

Theorem \ref{MT} establishes that, restricted to the ground state band, 
the full unitary group $U(t)$
and the approximate one-particle unitary $U_1(t)$ are uniformly close to each other and 
Theorems \ref{effop} and \ref{effop2} lift this assertion to semi-classical observables.
Experimentally measured are empirical statistics of suitable observables, like position and 
momentum, and we still have the task to investigate in what sense they are approximated
through the time evolution $U_1(t)$ for $\epsi \ll 1$.

For small $\epsi$, $H_1$ itself is a semi-classical Hamiltonian, which means that empirical 
distributions can be determined through the classical flow $\Phi^t$ generated by $H_1$.
Somewhat crudely the scheme is as follows: one choses an initial wave function
$\psi^{\epsi}$, which may or may not depend on $\epsi$, 
such that for small $\epsi$ it determines the measure $\rho_{\rm cl}(dx\,dp)$
on phase space. We evolve $\psi^\epsi$ as $\psi^\epsi_t = e^{-iH_1 t/\epsi}\psi^\epsi$ and
$\rho_{\rm cl}(dx\,dp,t) = (\rho_{\rm cl}\circ \Phi^{-t})(dx\,dp)$. Then the empirical
distributions computed from $\psi^\epsi_t$ agree with those of $\rho_{\rm cl}(t)$ up to
errors of order $\epsi$, i.e.\ quantum distributions are well approximated by their classical 
counterpart. In our context the true statistics must be compared with $U(t/\epsi)\psi^\epsi$
and we have to make sure that the approximations of Theorem \ref{effop2} are so sharp that
we can draw the same conclusions  for $U(t/\epsi)\psi^\epsi$. 

There are various ``schools'' which differ in what initial $\psi$'s are regarded as 
physically natural. 

\noindent 
(i) {\em  wave packet dynamics.} The initial wave function is well localized in  macroscopic 
position and  momentum, i.e.\ $\rho_{\rm cl}(dx\,dp) = \delta(x-x_0)\delta(p-p_0)\,dx\,dp$.
Then the wave packet follows the classical orbit , which only reflects that 
$\rho_{\rm cl}(dx\hspace{.5pt}dp,t) = \delta(x-x_t)\delta(p-p_t)\,dx\,dp$.

\noindent
(ii) {\em microscopic wave function independent of $\epsi$.} On the macrscopic scale
the position is localized, but there is momentum spread. Therefore $\rho_{\rm cl}
(dx\,dp) = \delta(x)|\widehat\psi(p)|^2\,dx\,dp$. Such a choice is appropriate immediately 
after a scattering event. Then $\psi$ is still localized at the scatterer but has considerable
momentum spread.

\noindent 
(iii) {\em WKB.} The wave function is taken 
to be build up from local plane waves, which means it has the
form $\psi^\epsi(x) = \epsi^{d/2} f(\epsi x) e^{iS(\epsi x)/\epsi}$ on the microscopic scale.
$\psi^\epsi$ is spread over macroscopic distance, but at any given point it has a sharp momentum.
$\psi^\epsi$ yields the phase space measure $\rho_{\rm cl}(dx\,dp) = |f(x)|^2 \delta(p-
\nabla S(x))\, dx \,dp$.

As will be discussed, all three classes of initial wave functions, and in principle more, 
can be handled under rather mild regularity assumptions in a unified fashion. The approximation
from $U(t/\epsi)$ to $U_1(t/\epsi)$ is covered by Theorems \ref{MT} and \ref{effop2} and
the semi-classical distributions for $U_1(t)$ are a consequence of standard results of
semi-classical analysis as presented, for example, in \cite{DS}. 
Since we did not find a discussion of sufficient generality in the literature,
we explain the arguments in some detail.

Note that  in the following we only treat the case $M=\IR^d$. If $M$ is a torus,
position and (quasi)-momentum are not related by Fourier transformation, but by a Bloch-Floquet
transformation. Apart from this difference the analysis goes through analogously.

\subsection{Wave packets following a classical trajectory}

The conceptually simplest way for a quantum particle to behave classically is to have a well
localized wave function that follows a classical trajectory.
Hence we consider initial wave functions with sharply peaked momentum and, at the same time, 
sharply peaked macroscopic position. Let the initial wave function be 
\[
\phi_{x_0,p_0}(x) = \epsi^{d/4}\, e^{ix \cdot p_0}\,\phi(\sqrt{\epsi}(x-x_0/\epsi))
\]
for some $\phi\in L^2(\IR^d)$, i.e., 
a wave function that is peaked on the macroscopic scale and centered at $x_0$, but spread out 
on the microscopic scale. Its Fourier transform is given by 
\[
\widehat \phi_{x_0,p_0} (p) = \epsi^{-d/4}\,e^{-i\frac{x_0\cdot( p-p_0)}{\epsi}} \,
\widehat\phi(\frac{p-p_0}{\sqrt{\epsi}})\,,
\]
which becomes sharply peaked at $p_0$ for $\epsi$ small. There is no difficulty to include also 
asymmetric scaling with weights $\epsi^{1-\alpha}$ and $\epsi^\alpha$, $0<\alpha<1$, 
and the choice $\alpha=1/2$ was made to simplify presentation.
Under the time evolution generated by 
\[
H_1 = E(-i\nabla_x) + V(\epsi x)
\]
it moves along the corresponding classical trajectory starting at $(x_0,p_0)$ following the classical
flow $\Phi^t$ generated by 
\[
H_{\rm cl} = E(p) + V(x)\,.
\]
To be consistent with the previous chapters, we continue to work in momentum representation.
\begin{mathe}{Proposition} \label{nospread}
Let $a\in S_0^0(1)$, $\widehat \phi \in H^1(\IR^d)\cap D(p)$ and $T<\infty$.  Then there is a $C<\infty$ such that  for 
 $t\in[0,T]$
\begin{equation}\label{wpe}
\left| \,\langle\widehat  \phi_{x_0,p_0},a_1^\epsi(t)\,\widehat \phi_{x_0,p_0}\rangle -
\big( a\circ\Phi^t\big)(x_0,p_0)\,\right|\leq\,C\sqrt{\epsi}\,\|\widehat \phi\|\big(\|\widehat \phi\|+\|p\widehat \phi\|+ 
\|\nabla_p\widehat \phi\|\big)\,.
\end{equation}
\end{mathe}

Using Theorem \ref{effop2}, this translates immediately to the full dynamics.

\begin{mathe}{Corollary} 
 Let the assumptions of Theorem \ref{effop2} be satisfied.
 Then there is a $C<\infty$ such that  for 
$\psi \in\big( (H^1(\IR^d)\cap D(p)) \otimes \mathcal{H}_{\rm f})\big)\,\cap$ Ran$P_{\rm i}$
and $t\in[0,T]$
\begin{eqnarray*}\lefteqn{
\left| \,\langle \psi_{x_0,p_0},a^\epsi(t)\,\psi_{x_0,p_0}\rangle -
\big( a\circ\Phi^t\big)(x_0,p_0)\,\right|}
\\ \\ &&\qquad
\leq\,C\sqrt{\epsi}\,\|\psi\|\big(\|\psi\|+
\|(p\otimes{\bf 1})\psi\|+ 
\|(\nabla_p\otimes{\bf 1})\psi\|\big)\,.
\end{eqnarray*}
\end{mathe}

Hence, when initially, on the macroscopic scale,  position and momentum are both sharply
defined, the wave packet follows the classical orbit without spreading even for 
macroscopic times. Such a situation occurs for example in particle accelerators, where
one can indeed calculate the particle trajectories based solely on classical dynamics
in good approximation.
\medskip

\noindent {\bf Proof} [of Proposition \ref{nospread}.] Referring to Egorov's theorem
we have to compute
\[
  \,\langle\widehat \phi_{x_0,p_0},(a\circ\Phi^t)^{\rm W,\epsi}\,\widehat\phi_{x_0,p_0}\rangle\,.
\]
One can replace $(a\circ\Phi^t)^{\rm W,\epsi}$ by the so called standard
quantization of $a\circ\Phi^t$ defined by
\begin{equation}
\big(
(a\circ\Phi^t)^{\rm S,\epsi}\, \widehat\phi\,\big)(p) := (2\pi)^{-d/2} \int dx \,
 \big(a\circ\Phi^t\big)
\left(\epsi x, 
p \right)\, e^{-ip\cdot x}\,\phi(x) 
\end{equation}
where the error is of order $\epsi$ uniformly in $t\in[0,T]$ (cf.\ Chapter 7 in \cite{DS}).
For the standard quantization the result becomes a simple calculation:
\begin{eqnarray*} \displaystyle\lefteqn{
 \,\langle\widehat \phi_{x_0,p_0},(a\circ\Phi^t)^{\rm S,\epsi}\,\widehat\phi_{x_0,p_0}\rangle \,= }\\ \\
& &= \, (2\pi)^{-d/2}
\int dx\,dp\, e^{i\frac{x_0\cdot (p-p_0)}{\epsi}}\, 
 \widehat\phi^*\left(\frac{p-p_0}{\sqrt{\epsi}}\right) \,
(a\circ\Phi^t)(\epsi x,p)\,e^{-ip\cdot x} \, e^{ix\cdot p_0} \,\times \\ \\ &&\hspace{7cm}
\phi\left(\sqrt{\epsi}\left(x-\frac{x_0}{\epsi}\right)\right)\,\\ \\
&&=\,  (2\pi)^{-d/2}
\int  dx\,dp\, \widehat\phi^*\left(\frac{p}{\sqrt{\epsi}}\right) \,
(a\circ\Phi^t)(\epsi x+x_0,p+p_0)\,e^{-ip\cdot x} \,
\phi\left(\sqrt{\epsi}\left(x\right)\right)\,\\ \\
&&=\, (2\pi)^{-d/2}
\int d\overline x\,d\overline p\, \widehat\phi^*(\overline p)(a\circ\Phi^t)(\sqrt{\epsi}\overline x+x_0,
\sqrt{\epsi}\overline p+p_0)\,e^{-i\overline{p}\cdot\overline  x} \,
\phi(\overline x)\,\\ \\
&&= \,
(a\circ\Phi^t)(x_0, p_0) \,+\, O(\sqrt{\epsi})\,,
\end{eqnarray*}
where the last equality follows from
a Taylor expansion of $a\circ\Phi^t$ around $(x_0,p_0)$ together with the assumption
that $\phi \in H^1(\IR^d)\cap D(p)$.
Note also that
$(\nabla_x (a\circ\Phi^t))\in S^0_0(1)$ and 
$(\nabla_p (a\circ\Phi^t))\in S^0_0(1)$
and hence $(\nabla_x (a\circ\Phi^t))^{\rm S,\epsi}$
, $(\nabla_p (a\circ\Phi^t))^{\rm S,\epsi}
\in
\mathcal{L}(L^2(\IR^d))$ (cf.\ Chapter 7 in \cite{DS})

\hfill $\qed$

\subsection{Initial wave function with momentum spread}

Generally wave functions are not of the special form described in the previous subsection.
Typically they arise from microscopic interactions and thus ``live'' on the microscopic scale
and do not depend on $\epsi$. But
if the shape of the initial wave function does not depend on $\epsi$ 
it will effectively look like a delta function on the macroscopic
scale at $t=0$. 
More precisely, let $\phi\in L^2(\IR^d)$, then the scaled
position distribution is $\epsi^{-d}|\phi(x/\epsi)|^2$ which converges to $\delta(x)$
as a measure. However, the scaled momentum distribution is still $|\widehat\phi(p)|^2$,
since in the quotient $x/t$ the $\epsi$'s cancel and thus the initially peaked
wave function will spread if evolved to times of order $\epsi^{-1}$.
More generally we consider as initial wave function
\[
\phi_{x_0}(x) = \phi(x-x_0/\epsi)\,,
\]  
i.e.\ we move $\phi$ to the macroscopic initial position $x_0$.

Then it is natural to chose
\[
\rho_{\rm cl}(dx\,dp) = \delta(x-x_0)|\widehat\phi_{x_0}(p)|^2\,dx\,dp
\] 
as the corresponding classical phase space distribution at $t=0$ and evolve it according
to the classical flow to 
\[
\rho_{\rm cl}(dx\,dp,t) = \left(\rho \circ \Phi^{-t}\right) (dx\,dp)\,.
\]

Let us now, as the simplest example, compare the quantum mechanical position distribution
\[
\rho_{\epsi} (dx,t) = \epsi^{-d}|\phi_t(x/\epsi)|^2 \,dx\,,
\qquad \phi_t = e^{-iH_1t/\epsi}\phi_{x_0}\,,
\]
with the classical one,
\[
\rho^x_{\rm cl}(dx,t) = \int \rho_{\rm cl}(dx\,dp,t)\,,
\]
in the limit $\epsi\to 0$. As a first step we calculate, with $f\in C_0^\infty$ 
a test function,
\begin{eqnarray}
\int \rho_{\epsi} (dx,t) \,f(x)& = &\langle \phi^\epsi_t, f \,\phi^\epsi_t \rangle =
 \langle \phi_{x_0}, f(x^\epsi(t))\,\phi_{x_0}\rangle \nonumber \\ \label{rof}
&=&  \langle \widehat\phi_{x_0}, (f\circ \Phi^t_x)^{\rm W,\epsi}\,\widehat\phi_{x_0}\rangle +O(\epsi^2)
\,.
\end{eqnarray}
For the last equality we used Egorov's Theorem.

To proceed we need to know how the Weyl quantization of a time evolved classical observable
acts on microscopic wave functions.
We postpone the proof of the following Proposition to the end of the section.
\begin{mathe}{Proposition} \label{WeylAct}
Let $a\in S^0_0(1)$. Then for each $T<\infty$ there is a $C<\infty$ such that for
 $\phi\in D(x)$ and $t\in[0,T]$
\begin{equation} \label{WAE}
\Big\|
\Big(
\big(a\circ\Phi^t\big)^{\rm W,\epsi}
 - \big( a\circ\Phi^t\big)(x_0,\cdot)  \Big) \,
\widehat \phi_{x_0}\Big\| \leq \,C\,\epsi\,\|x\phi\|\,,
\end{equation}
where $ \big( a\circ\Phi^t\big)(x_0,\cdot)$ denotes the operator of multiplication
with the function $ \big( a\circ\Phi^t\big)(x_0,p)$ in momentum representation.
\end{mathe}

Thus (\ref{rof}) becomes
\begin{equation}\label{rof2}
 \langle\widehat \phi_{x_0}, (f\circ \Phi^t_x)^{\rm W,\epsi}\,\widehat\phi_{x_0}\rangle  = 
\int dp\, |\widehat \phi_{x_0}(p)|^2\, \big(f\circ\Phi_x^t\big)(x_0,p) +O(\epsi)\,.
\end{equation}
However, the right hand side of (\ref{rof2}) is exactly what we were aiming for:
\begin{eqnarray*}
\int dp\, |\widehat \phi_{x_0}(p)|^2\, \big(f\circ\Phi_x^t\big)(x_0,p) &=&
 \int \rho_{\rm cl}(dx\,dp,0) \big(f\circ\Phi_x^t\big)(x,p) \\
& = & \int \rho_{\rm cl}(dx\,dp,t) f(x)\\
& = & \int   \rho^x_{\rm cl}(dx,t) f(x) \,.
\end{eqnarray*}

In summary, one obtains that
\begin{equation}\label{rhoweak}
\lim_{\epsi\to 0} \rho_{\epsi} (dx,t) =  \rho^x_{\rm cl}(dx,t) 
\end{equation}
weakly as measures. Note that this means that the position distribution of the quantum particle
converges to the classical one, although the wave function is {\em not} following the classical
orbit, but spreading. Such a situation occurs for example in a scattering experiment. 
After the quantum particle is scattered off the target its wave function usually has a 
large momentum spread and thus it also spreads in position space. 
However, during the process of detection it is subject to relatively weak potentials and
can be treated like a classical particle, at least on the level of statistics.

Such a result holds  in fact not only for the position distribution but  
for all semi-classical observables.
For general observables $a\in S^0_0(1)$ we get analogously 
\begin{equation} \label{grof}
 \langle\widehat \phi_{x_0}, a_1^\epsi(t)\,\widehat\phi_{x_0}\rangle
=  \int \rho_{\rm cl}(dx\,dp,t) a(x,p)\,+\,O(\epsi)\,.
\end{equation}

Now Theorem \ref{effop2} allows us to immediately 
translate this result to the full quantum dynamics.
\begin{mathe}{Corollary} \label{spreadcor}
 Let the assumptions of Theorem \ref{effop2} be satisfied.
 Then there is a $C<\infty$ such that for 
$\psi \in (H^1(\IR^d)\otimes \mathcal{H}_{\rm f})\,\cap$ Ran$P_{\rm i}$ and $t\in[0,T]$
\[
\left|
 \,\langle \psi_{x_0},a^\epsi(t)\,\psi_{x_0}\rangle -  \int \rho_{\rm cl}(dx\,dp,t) a(x,p)
\right|<\,C\epsi\,\|\psi\|(\|\psi\|+\|(\nabla_p\otimes{\bf 1})\psi\|)\,.
\]
\end{mathe}

Before we close the section with the proof of Lemma \ref{WeylAct}, let us shortly 
comment on an approach ``dual'' to Weyl quantization of classical observables.
One can translate the above result  to the level of distributions, 
i.e.\ generalize (\ref{rhoweak}), by looking at the
scaled Wigner distribution. For $\phi\in L^2(\IR^d)$ it is defined through
\[
W^\epsi(dx\,dp,t) = \int d\xi\,
 \epsi^{-d} \,\phi_{x_0}^*(x/\epsi- \xi/2,t)\,\phi_{x_0}(x/\epsi +\xi/2,t)\,
e^{ip\cdot \xi}\,dx\,dp\,.
\]
The scaled Wigner distribution yields expectations of Weyl quantized operators through 
\[
\langle\widehat \phi_{x_0},  \,a_1^{\epsi}(t)\,\widehat\phi_{x_0} \rangle
= \int \, W^\epsi(dx\,dp,t)\,a(x,p)\,.
\]
This together with  (\ref{grof}) implies
\[
\lim_{\epsi\to 0} W^\epsi(dx\,dp,t) = \rho_{\rm cl}(dx\,dp,t)
\]
weakly as measures. 

On the level of the  full quantum dynamics we introduce
 the reduced Wigner distribution  for 
$\psi\in L^2(\IR^d)\otimes \mathcal{H}_{\rm f}$ as
\[
W_{\rm rd}^\epsi(dx\,dp,t) = \int d\xi\, \epsi^{-d} \,\big\langle \widehat \psi_{x_0}^*(x/\epsi- \xi/2,t),
\widehat \psi_{x_0}(x/\epsi +\xi/2,t)
\big\rangle_{\mathcal{H}_{\rm f}}\,
e^{ip\cdot \xi}\,dx\,dp\,,
\]
where here $\,\widehat{}\,$ stands for Fourier transformation in the first argument, i.e.\ for
$\mathcal F\otimes {\bf 1}$. Corollary \ref{spreadcor} translates into
\[
\lim_{\epsi\to 0}W_{\rm rd}^\epsi(dx\,dp,t) =  \rho_{\rm cl}(dx\,dp,t)
\]
weakly as measures.

\noindent {\bf Proof} [of Proposition \ref{WeylAct}].\quad
Again we replace $(a\circ\Phi^t)^{\rm W,\epsi}$ by the standard
quantization of $a\circ\Phi^t$ and get
\begin{eqnarray}
\lefteqn{\hspace{-.8cm}\big(
(a\circ\Phi^t)^{\rm S,\epsi}\, \widehat\phi_{x_0}\,\big)(p) := (2\pi)^{-d/2} \int dx \,
 \big(a\circ\Phi^t\big)
\left(\epsi x, 
p \right)\, e^{-ip\cdot x}\,\phi(x-x_0/\epsi)}\nonumber\\
 &=& (2\pi)^{-d/2} \int dx \,\big(a\circ\Phi^t\big)
\left(\epsi x+x_0, 
p \right)\, e^{-ip\cdot x}\,e^{-ip\cdot x_0/\epsi}\,\phi(x) \,.\label{sq}
\end{eqnarray}
Taylor expansion yields 
\[
(a\circ\Phi^t)(\epsi x+x_0,p) = (a\circ\Phi^t)(x_0,p) + \epsi x\cdot(\nabla_x (a\circ\Phi^t))(y(x),p)\,.
\]
We insert this into (\ref{sq}) and conclude as in the proof of Proposition \ref{nospread} that 
the term proportional to $\epsi$ is bounded in 
norm by our assumption that
$\phi\in D(x)$. This proves (\ref{WAE}).

\hfill$\qed$

\subsection{Initial wave function of WKB form}

In the previous case of an initially localized wave function the different momentum components
travel at different velocities and therefore such a wave function spreads on the macroscopic
scale. After some time one expects the wave function to have locally well defined momentum as long
as no interference occurs, i.e.\ it should be of WKB type.
On the microscopic scale a WKB wave function has the form
\[
\phi(x) = \epsi^{d/2} f(\epsi x) e^{i\frac{S(\epsi x)}{\epsi}}\,,
\]
with $f$ and $S$ real valued.
Hence it locally  looks  like a plane wave with momentum 
$\nabla S(\epsi x)$ and amplitude $f(\epsi x)$. 

Time-dependent WKB approximation is concerned with showing that the time evolution of
such a wave function is in first order given by 
\[
\big(e^{-iH_1 t/\epsi}\phi\big) (x) \approx f_t(\epsi x) e^{i\frac{S_t(\epsi x)}{\epsi}}\,,
\]
where $S_t$ is the solution of the classical Hamilton-Jacobi equation with initial condition
$S$ and $f_t$ is the solution of the classical continuity equation
\[
\partial_t f + {\rm div} \nabla S_t = 0\,.
\] 
The corresponding classical phase space distribution is therefore
\begin{equation}\label{WKBdist}
\rho_{\rm cl} (dx\,dp) = f^2(x) \delta( p- \nabla S(x))\,dx\,dp\,.
\end{equation}
The main drawback of the WKB approximation is that it works only as long as no 
caustics are reached, 
or, put differently, as long as no interference between different parts of 
the WKB wave function happens.

However, if we focus again on distributions of semi-classical observables
on phase space no difficulty arises, only $\rho_{\rm cl}(t)$ is no longer of the particular form 
(\ref{WKBdist}).

\begin{mathe}{Proposition} \label{WKBprop}
Let $f\in C^\infty(\IR^d)\cap L^1(\IR^d)$, $S\in C^\infty(\IR^d)$ and
\[
\phi(x) = \epsi^{d/2} f(\epsi x)  e^{i\frac{S(\epsi x)}{\epsi}}\,
\]
be normalized in $L^2(\IR^d)$.
Let  $a\in C^\infty_0(\IR^d\times \IR^d)$ and $T<\infty$. Then there is a $C<\infty$ such that
for all $t\in [0,T]$
\begin{equation}
\left|
\langle\widehat \phi, a_1^\epsi(t)\widehat  \phi \rangle - 
\int  \rho_{\rm cl}(dx\,dp, t) a(x,p)
\right|  
\leq \,C\sqrt{\epsi}\,,
\end{equation}
where $ \rho_{\rm cl}(dx\,dp, t) = (\rho_{\rm cl}\circ \Phi^{-t})(dx\,dp)$
and $\rho_{\rm cl}(dx\,dp)$ was defined in (\ref{WKBdist}).
\end{mathe}

As in the preceding cases we can again translate Proposition \ref{WKBprop} to the full dynamics
using Theorem \ref{effop2}, but we omit the corresponding statement this time.

\noindent {\bf Proof.}\quad
We apply Egorov and switch to standard quantization:
\[
\langle\widehat \phi, a_1^\epsi(t) \widehat  \phi \rangle = \langle\widehat  \phi, (a\circ\Phi^t)^{\rm W,\epsi} \widehat  \phi \rangle
+ O(\epsi)
= 
\langle\widehat  \phi, (a\circ\Phi^t)^{\rm S,\epsi} \widehat  \phi \rangle
+ O(\epsi)\,.
\]
We calculate for $a\in C^\infty_0$
\begin{eqnarray*}
\widehat{ (a^{\rm S,\epsi}\widehat\phi\,)}\big(\frac{y}{\epsi}\big) &=& 
(2\pi)^{-d} \epsi^{d/2} \int dx\,dp\, a(\epsi x,p) f(\epsi x) e^{-ip\cdot(x-y/\epsi)} e^{iS(\epsi x)/\epsi}
\\
&=&
(2\pi)^{-d} \epsi^{-d/2} \int dx\,dp\, a( x,p) f( x) e^{i(S(x)- p\cdot(x- y))/\epsi}
\\
&=&
(2\pi)^{-d} \epsi^{-d/2} \int dz\,dp\, a( z+y,p) f( z+y) e^{i(S(z+y)- p\cdot z)/\epsi}\,.
\end{eqnarray*}
Stationary phase method (cf.\ Theorem 7.7.7 in \cite{Hoermander} with $k=n+1$) yields that
\begin{eqnarray*}
\left|\int dz\,dp\, a( z+y,p) f( z+y) e^{i(S(z+y)- p\cdot z)/\epsi} -\right.\\ & \left.\hspace{-4cm}
(2\pi\epsi)^d a(y,\nabla S(y))f(y) e^{\frac{i}{\epsi} S(y)} \right| \leq  C\,\epsi^{d+\frac{1}{2}}\,,
\end{eqnarray*}
where the constant is uniform in $y$ and depends on $a$ via a sum of sup-norms of finite many partial
derivatives. Hence
\[
\widehat{((a\circ\Phi^t)^{\rm S,\epsi} \widehat  \phi)}\big(\frac{y}{\epsi}\big) = \epsi^{d/2} 
(a\circ\Phi^t)(y,\nabla S(y))f(y) e^{\frac{i}{\epsi} S(y)} + \epsi^{d/2}O(\sqrt{\epsi})
\]
and 
\[
\langle\widehat  \phi, (a\circ\Phi^t)^{\rm S,\epsi} \widehat  \phi \rangle = \int dy\, 
(a\circ\Phi^t)(y,\nabla S(y)) f^2(y) + O(\sqrt{\epsi})\,,
\]
where we used that $f\in L^1(\IR^d)$.

\hfill$\qed$

\end{document}